\documentclass[letterpaper, 10 pt, conference]{ieeeconf}  

\IEEEoverridecommandlockouts                              

\usepackage{graphicx}%
\usepackage{multirow}%
\usepackage{amsmath,amssymb,amsfonts}%
\usepackage{mathrsfs}%
\usepackage{textcomp}
\usepackage{xcolor}%
\usepackage{textcomp}%
\usepackage{manyfoot}%
\usepackage{booktabs}%
\usepackage{algorithm}%
\usepackage{algorithmicx}%
\usepackage[noend]{algpseudocode}%
\usepackage{listings}%
\usepackage{caption}
\usepackage{subcaption}
\usepackage{svg}
\usepackage{stfloats}
\usepackage{float}
\usepackage{placeins}
\usepackage{cite}
\usepackage{hyperref}
\usepackage{tabularx} 
\usepackage{siunitx}
\usepackage{makecell}
\usepackage{adjustbox}
\usepackage{url}
\usepackage{bm}

\algnewcommand{\algorithmicgoto}{\textbf{go to}}%
\algnewcommand{\Goto}[1]{\algorithmicgoto~\ref{#1}}%

\title{\LARGE \bf
Collaborative Task Assignment, Sequencing and Multi-agent Path-finding
}

\author{Yifan Bai, Shruti Kotpalliwar, Christoforos Kanellakis and George Nikolakopoulos%
\thanks{The authors are with the Robotics and AI group, in the Department of Computer Science, Electrical and Space Engineering at Luleå University of Technology, Sweden.}
\thanks{This work has been partially funded by the European Union’s Horizon
Europe Research and Innovation Programme under the Grant Agreement No.101138451 PERSEPHONE.}
}

\begin{document}
\newcommand{\customcomment}[1]{ // #1}
\maketitle

\begin{abstract} In this article, we address the problem of collaborative task assignment, sequencing, and multi-agent pathfinding (TSPF), where a team of agents must visit a set of task locations without collisions while minimizing flowtime. TSPF incorporates agent-task compatibility constraints and ensures that all tasks are completed. We propose a Conflict-Based Search with Task Sequencing (CBS-TS), an optimal and complete algorithm that alternates between finding new task sequences and resolving conflicts in the paths of current sequences. CBS-TS uses a mixed-integer linear program (MILP) to optimize task sequencing and employs Conflict-Based Search (CBS) with Multi-Label A* (MLA*) for collision-free path planning within a search forest. By invoking MILP for the next-best sequence only when needed, CBS-TS efficiently limits the search space, enhancing computational efficiency while maintaining optimality.

We compare the performance of our CBS-TS against Conflict-based Steiner Search (CBSS), a baseline method that, with minor modifications, can address the TSPF problem. Experimental results demonstrate that CBS-TS outperforms CBSS in most testing scenarios, achieving higher success rates and consistently optimal solutions, whereas CBSS achieves near-optimal solutions in some cases. The supplementary video is available at \url{https://youtu.be/QT8BYgvefmU}.
\end{abstract}

\section{Introduction}
\begin{figure}[t]
    \centering
    \begin{subfigure}{.48\linewidth}
    \centering
    \includegraphics[width=0.85\linewidth]{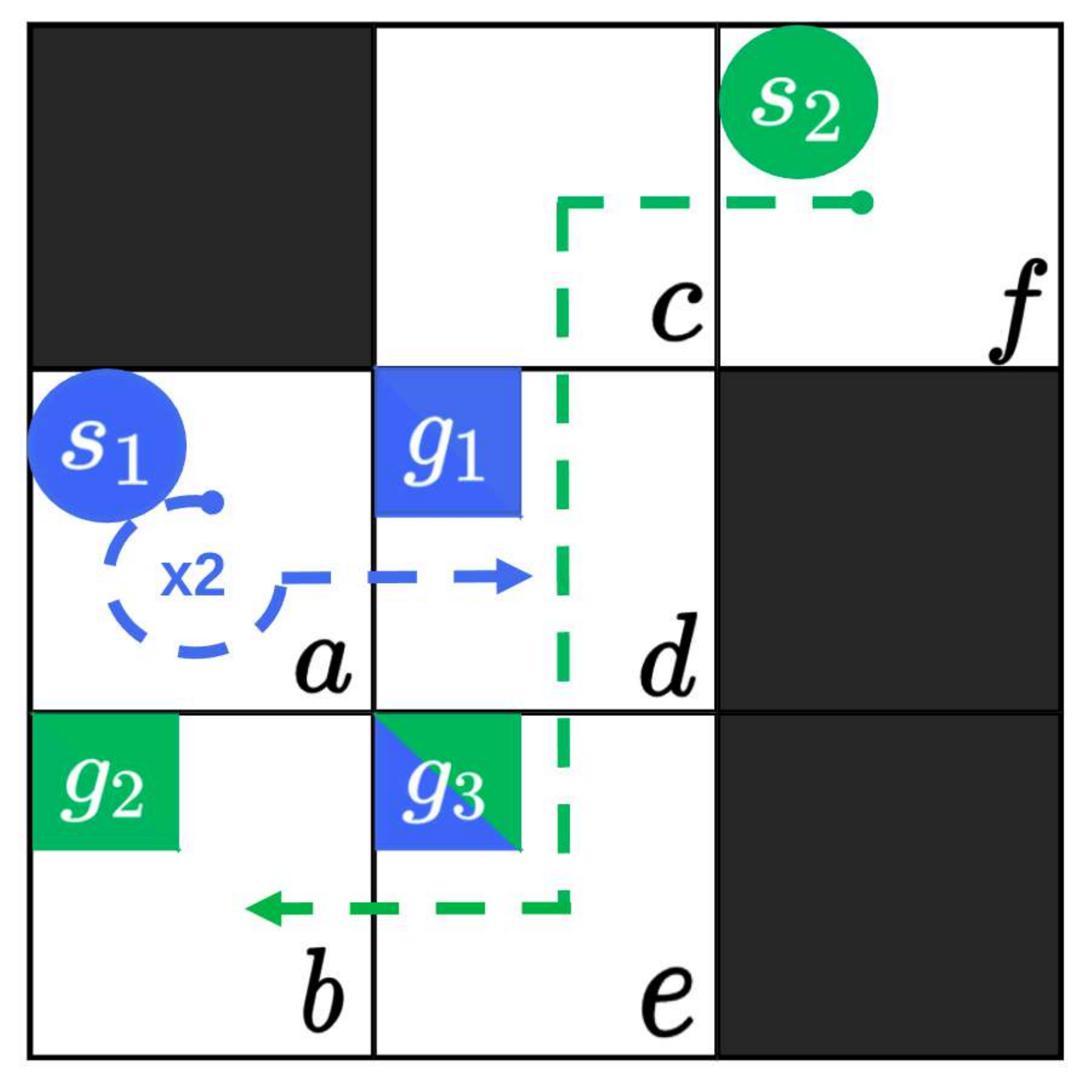}
    \caption{}
    \label{fig:concept1}
    \end{subfigure}
    \begin{subfigure}[b]{.48\linewidth}
    \centering
    \includegraphics[width=0.85\linewidth]{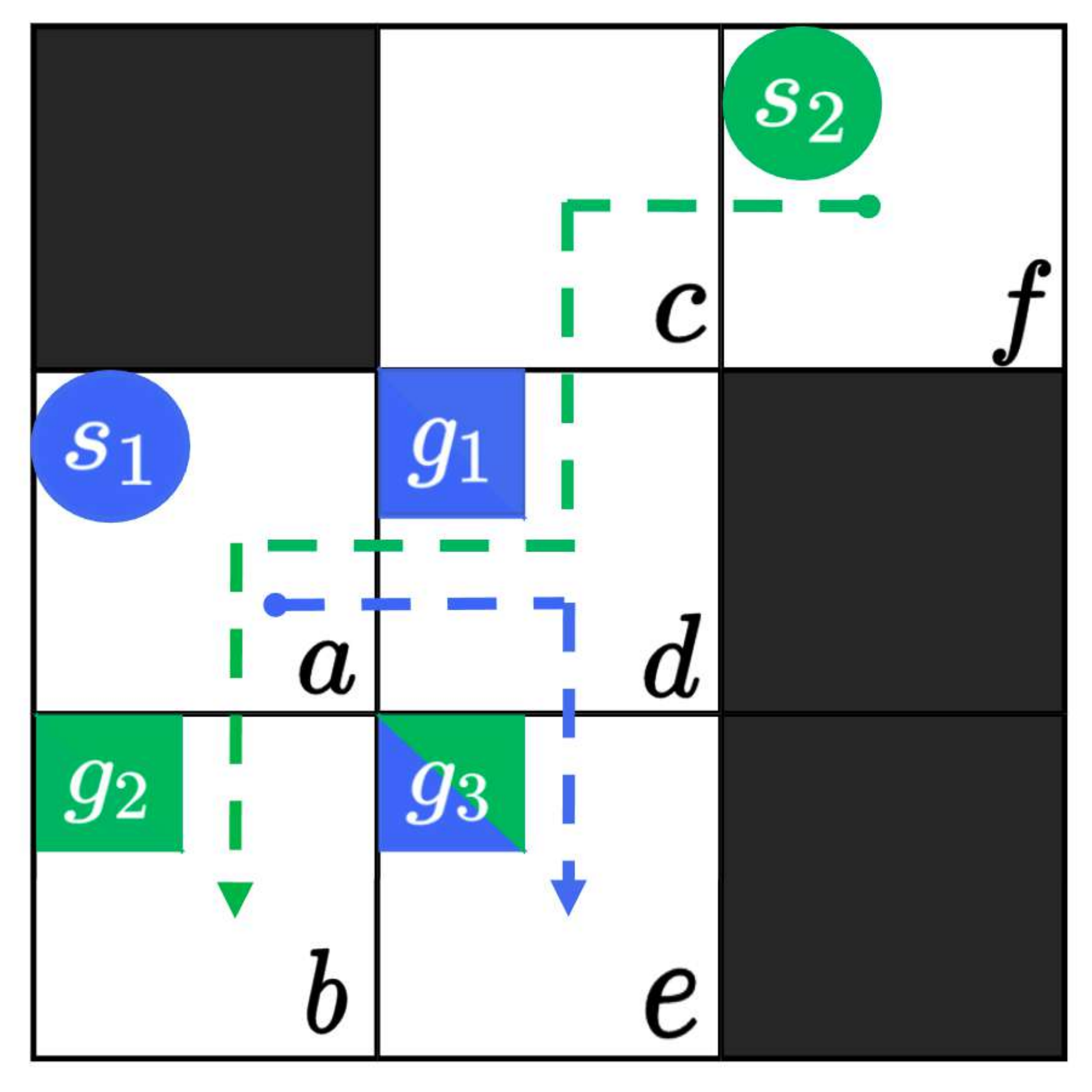}
    \caption{}
     \label{fig:concept2}
    \end{subfigure}
    \caption{An example demonstrating how different task assignments result in different conflict-free paths for two agents (starting at $s_1$, $s_2$) and three tasks (at $g_1$, $g_2$, $g_3$), with dashed lines representing the planned conflict-free paths. The circular segment in the blue path represents a wait action required to resolve conflicts. The black cells denote obstacles. The color of the starts and goals
    indicates the agent-task compatibility constraints (i.e. $g_3$ can be visited by either the green or blue agent).}
    \label{fig:concept}
\end{figure}

Deploying multiple autonomous robots in domains such as warehouse pick-up and delivery~\cite{ma2019lifelong} or search and rescue operations~\cite{drew2021multi} involves addressing several challenges, among which task assignment\cite{chakraa2023optimization} and multi-agent pathfinding (MAPF)\cite{stern2019multi} play a critical role. Task assignment determines which agents should handle which tasks, while MAPF ensures agents navigate from their start locations to task destinations without collisions. If the objective is to achieve overall optimality (e.g., minimizing flowtime), these two problems are tightly coupled, as certain task assignments can lead to fewer collisions during path planning. For instance, Figure~\ref{fig:concept1} illustrates a task sequencing that initially minimizes flowtime to 5 without accounting for potential collisions. In this sequence, agent 1 is assigned tasks $g_1$, and agent 2 is assigned $g_3$ followed by $g_2$. However, resolving conflicts in the resulting paths requires agent 1 to wait at cell $a$ for 2 time steps, increasing the total flowtime to 7. In contrast, Figure~\ref{fig:concept2} presents an alternative sequencing result with a higher initial flowtime of 6, but it remains conflict-free, leading to a better overall solution. This demonstrates that task assignment and path planning must be jointly considered to achieve optimal execution.

In this article, we address the collaborative task sequencing and pathfinding (TSPF) problem and propose an optimal solution that minimizes flowtime while ensuring all tasks are completed and agent-task compatibility constraints are satisfied. 

\subsection{Related Work}
\textbf{Task Assignment:}
Chakraa et al.\cite{chakraa2023optimization} provided a comprehensive taxonomy of multi-robot task assignment problems. Related problems include the Multiple Traveling Salesman Problem (mTSP)\cite{cheikhrouhou2021comprehensive}, where agents must visit a set of locations in an optimal sequence, and the Vehicle Routing Problem (VRP)\cite{konstantakopoulos2022vehicle}, a variant of mTSP with additional constraints. A wide range of mTSP solvers has been developed, including exact algorithms\cite{sundar2015exact, benavent2013multi} for optimal solutions and heuristic methods~\cite{karabulut2021modeling, soylu2015general} for near-optimal solutions. TSPF reduces to mTSP when salesmen start from different depots (i.e., agent start positions), are not required to return to their depots, and conflicts between salesmen are ignored.

\textbf{Multi-agent path finding (MAPF):} A variety of methods have been proposed to solve MAPF~\cite{wagner2011m, de2014push, yu2016optimal}. Among them, Conflict-Based Search (CBS)\cite{sharon2015conflict} has gained prominence due to its optimality and completeness. However, solving MAPF optimally is NP-hard\cite{yu2015optimal}, motivating research into enhancements~\cite{lazyCBS, symmetryreasoning, boyarski2015icbs} that improve CBS’s efficiency. Additionally, several suboptimal MAPF solvers~\cite{ma2019searching, barer2014suboptimal} have been developed to achieve faster runtimes. Since our method builds on CBS, these optimizations and suboptimal variants can also be applied to improve its efficiency.

\textbf{Combined task assignment and path-finding (TAPF):} TAPF\cite{ma2016TAPF} extends MAPF by integrating task allocation, where the number of single-goal tasks matches the number of agents, and each agent is assigned exactly one task. One leading approach CBS-TA~\cite{honig2018conflict} guarantees optimality through a search forest but assigns only one goal per agent, leaving some goals unvisited. Extensions like~\cite{zhong2022optimal} model tasks as sequences of fixed-ordered goals, ensuring complete goal assignment but preventing goals within a task from being assigned to different agents. Multi-Goal MAPF (MG-MAPF)~\cite{surynek2021multi} does not perform task assignment; instead, each agent is given a predefined set of goals to visit. While MG-MAPF allows agents to visit their assigned goals in arbitrary order and guarantees optimality, it does not permit goals to be assigned across different agents. The Multi-Agent Pickup and Delivery (MAPD)\cite{ma2019lifelong} problem addresses cases where agents must complete tasks that involve visiting two distinct locations, such as picking up and delivering an item. Henkel et al.\cite{henkel2019optimal} solve MAPD optimally via brute-force enumeration, but their paper only reports results for up to four tasks, leaving its scalability unclear. PC-TAPF\cite{brown2020optimal} incorporates precedence constraints between tasks, but its low-level path planner lacks completeness guarantees, hindering robustness in complex scenarios. Recent work, CBSS~\cite{ren2023cbss}, supports full goal coverage and allows any goal to be assigned to any agent while being theoretically optimal. However, its implementation remains computationally inefficient, even with a heuristic solver that does not guarantee optimality.

While existing methods tackle various aspects of task assignment and pathfinding, they face critical limitations, such as incomplete goal coverage\cite{honig2018conflict,tang2023solving}, rigid goal assignment within predefined sets\cite{zhong2022optimal,surynek2021multi}, suboptimal\cite{sa-recbs, ren2024bounded} or incomplete\cite{brown2020optimal} solutions, and computational inefficiencies\cite{henkel2019optimal, ren2023cbss}. These limitations highlight the need for a scalable and optimal approach for TSPF, which guarantees all goals are visited while allowing goals to be assigned freely, subject to agent-task constraints.
\subsection{Contribution}
In this paper, we propose CBS-TS, an efficient, optimal, and complete approach to the TSPF problem. CBS-TS ensures that all goal locations are visited while allowing any goal to be assigned to any agent, provided the agent-task constraints permit it. 

CBS-TS employs a two-level hierarchical approach. At the high level, the algorithm constructs a search forest by iteratively alternating between two processes: (1) identifying the next-best task sequence (root node of a new search tree) and (2) resolving conflicts to generate collision-free paths for the given task sequence (exploring the solution in a search tree). At the low level, CBS-TS leverages Multi-label A* (MLA*) \cite{grenouilleau2019multi} to compute collision-free paths for agents to visit their assigned tasks in the specified sequence. 

A key contribution of this work is the introduction of a Mixed-Integer Linear Programming (MILP) formulation to efficiently determine the next-best task sequence. We provide theoretical proof demonstrating that CBS-TS is both optimal and complete. 

Through extensive experiments, we compare CBS-TS with the CBSS algorithm, demonstrating its better performance in terms of effectiveness and efficiency. Additionally, we validate the practicality of our approach through physical experiments, further underscoring its applicability in real-world scenarios.

\section{Problem Description } \label{problemDescription}
Consider an undirected graph $G=\left(V, E \right)$, where $V$ represents a set of locations and $E \subseteq V \times V$ represents connections between these locations. We are given a set of $N$ heterogeneous agents $\mathcal{A}=\left\{a_1, a_2, \ldots, a_N\right\}$, each starting from an initial location in $\mathcal{S}=\left\{s_1, s_2, \ldots, s_N\right\} \subseteq V$, where $s_i$ denotes the initial location of agent $a_i$. Additionally, there are $M$ goal locations, $\mathcal{G}=\left\{g_1, g_2, \ldots, g_M\right\}\subseteq V$, each of which must be visited by an agent. In this work, tasks refer to the requirement for agents to visit specific goal locations in the environment.

To model the agent-task compatibility constraints, we define a binary compatibility matrix $\Delta$ of size $N \times M$, where each entry $\delta_{ij} \in \{0,1\}$ indicates whether agent $a_i$ is permitted to reach goal $g_j$. Specifically, $\delta_{ij}=1$ if agent $a_i$ is allowed to reach goal $g_j$, and $\delta_{ij}=0$ otherwise.

All agents operate under a global clock. At each discrete time step, an agent can either wait at its current vertex or move to an adjacent vertex. The objective is to determine a task sequencing and corresponding paths $\pi_i=\left(v_i^0, v_i^1, \ldots, v_i^{T_i}\right)$ for each agent $a_i$, such that the following conditions are satisfied:
\begin{enumerate}[noitemsep]
    \item \textbf{Initial Condition: } Each agent $a_i$ starts at its designated initial location, i.e. $v_i^0=s_i, s_i \in \mathcal{S}$;
    \item \textbf{Terminal Condition: } If agent $a_i$ assigned tasks, all goal locations in its assigned task sequence must satisfy the agent-task compatibility constraints, and $a_i$ stops at the last goal location in that task sequence and remains there, i.e. $v_i^{T_i}=g_j,  g_j \in \mathcal{G}, \delta_{i j}=1, v_i^t=g_j, \forall t>T_i$ \\
    If not assigned any tasks, agent $a_i$ stays at its starting location, i.e. $v_i^t=s_i, \forall t$;
    \item \textbf{Goal Coverage: } Each goal location $g_j \in \mathcal{G}$ must be visited exactly once by an agent $a_i$, provided $\delta_{ij}=1$, i.e. $\forall g_j \in \mathcal{G}, \quad \exists i, t: v_i^t=g_j \wedge \delta_{i j}=1$;
    \item \textbf{Conflict-Free Paths: }The paths of all agents must be conflict-free, ensuring:\\$v_i^t \neq v_j^t, \forall t, \forall i \neq j$ (no vertex conflict)\\
    $\left(v_i^t, v_i^{t+1}\right) \neq\left(v_j^{t+1}, v_j^t\right), \forall t,\forall i \neq j$ (no edge conflict);
    \item \textbf{Objective: } The flowtime $\sum_{i=1}^N T_i$ is minimized. 
\end{enumerate}  

\noindent\textbf{Remark: }Our method can be easily adapted to minimize the makespan $\max_{1 \leq i \leq N} T_i$ by modifying the objective function in the MILP formulation. However, in this paper, we focus solely on minimizing flowtime to ensure a fair comparison with the baseline method CBSS under the same metric.
\section{Methodology}
In this section, we propose CBS-TS (Conflict-Based Search with Task Sequencing), a two level method inspired from CBS-TA\cite{honig2018conflict} but designed to solve TSPF problems. At the high level, CBS-TS extends CBS framework by incorporating optimal task sequencing through a Mixed-Integer Linear Programming (MILP) model, ensuring all goal locations are visited. At the low level, CBS-TS utilizes MLA* to plan collision-free paths for agents to visit their specific task sequences.

\subsection{High Level of CBS-TS}~\label{conflictforest}
CBS-TS builds upon CBS\cite{sharon2015conflict}, which uses a binary tree to resolve conflicts and find collision-free paths for MAPF. CBS-TS incorporates task sequencing through a search forest, where each tree corresponds to the CBS conflict resolution of a distinct task sequencing. Each node in the forest consists of: (1) A boolean flag \textit{root} that indicates whether the node is a root node, (2) The task \textit{sequencing} associated with the tree, (3) A set of \textit{constraints}, where vertex constraint $\left\langle a_i, u, t\right\rangle$ prevents agent $a_i$ from occupying location $u$ at time $t$ and edge constraint $\left\langle a_i, u, v, t\right\rangle$ prohibits agent $a_i$ from moving from $u$ to $v$ at time $t$, (4) A set of \textit{paths} for all agents that satisfy the constraints, and (5) \textit{Cost}, measured as the total flowtime of all paths. 

The pseudocode for CBS-TS is provided in Algorithm~\ref{alg:highlevel}. The algorithm starts by generating the root node of the first tree. This node contains the best task sequencing, solved by MILP (Section~\ref{MILP}), while ignoring conflicts among agents. Initial paths for each agent are then computed using MLA* based on the assigned task sequencing, without any vertex/edge constraints. This root node is added to the OPEN list [Lines~\ref{rootstart}–\ref{rootend}]. During execution, the algorithm iteratively selects the lowest cost node $P$ from the OPEN list [\textbf{Line}~\ref{bestnode}]. If paths in $P$ are conflict-free, the algorithm returns the solution [\textbf{Lines}~\ref{solutionfirst}-\ref{solutionend}]. Otherwise, the earliest conflict (vertex or edge) is identified and resolved by generating two child nodes, each inheriting the parent's constraints and adding a new constraint derived from the conflict. The path of the constrained agent is replanned using MLA* [\textbf{Lines}~\ref{conflictstart}-\ref{conflictend}].

To explore alternative task sequencings, CBS-TS creates a new root node $R^{\prime}$ with the next best task sequencing before expanding the child nodes of the current root node [\textbf{Lines}~\ref{newrootstart}-\ref{newrootend}]. The next-best sequencing is determined by the MILP solver while excluding previously found sequencings in $\mathcal{P}_\text{set}$ [\textbf{Line}~\ref{nextbest}]. By conducting the best-first search across the entire search forest, it is ensured that the globally optimal flowtime solution is found among all possible task sequencing options. 
\begin{algorithm}[ht]
\small
\caption{High Level of CBS-TS}\label{alg:highlevel}
\begin{algorithmic}[1]
\Statex \textbf{Input:} $N$, $M$, cost matrix $C$, $\Delta$
\Statex \textbf{Output:} optimal path for each agent
\State $\mathcal{P}_{\text{set}} \leftarrow \emptyset$ \customcomment{previously found solution set}\label{rootstart}
\State $R.\text{root} \leftarrow$ \textit{true} 
\State $R.\text{sequencing} \leftarrow \text{MILP}(N,M,C,\Delta, \mathcal{P}_{\text{set}})$
\State Insert $R.\text{sequencing}$ to $\mathcal{P}_{\text{set}}$
\State $R.\text{constraints} \leftarrow \emptyset$
\For{$i \gets 1$ to $N$}
    \State $R.\text{paths}[a_i] \leftarrow \text{MLA}^*(a_i, R.\text{sequencing}[a_i], R.\text{constraints})$
\EndFor
\State $R.\text{cost} \leftarrow \text{flowtime}(R.\text{paths})$
\State Insert $R$ to OPEN\label{rootend}
\While{OPEN $\neq \emptyset$}
    \State $P \leftarrow \text{best node from OPEN}$\customcomment{lowest solution cost} \label{bestnode}
    \If{$P.\text{paths do not have collisions}$}\label{solutionfirst}
        \State \textbf{Return} $P.\text{paths}$
    \EndIf\label{solutionend}
    \If{$P.\text{root is \textit{true}}$}\label{newrootstart}
        \State $R^{\prime}.\text{root} \leftarrow$ \textit{true}  \customcomment{new root node (next best sequencing)}
        \State $R^{\prime}.\text{sequencing} \leftarrow \text{MILP($N,M,C,\Delta,\mathcal{P}_{\text{set}}$)}$\label{nextbest}
        \State Insert $R^{\prime}.\text{sequencing}$ to $\mathcal{P}_{\text{set}}$
        \State $R^{\prime}.\text{constraints} \leftarrow \emptyset$
        \For{$i \gets 1$ to $N$}
        \scriptsize
        \State $R^{\prime}.\text{paths}[a_i] \leftarrow \text{MLA}^*(a_i, R^{\prime}.\text{sequencing}[a_i], R^{\prime}.\text{constraints})$
        \small
        \EndFor
        \State $R^{\prime}.\text{cost} \leftarrow \text{flowtime}(R^{\prime}.\text{paths})$
        \State $R^{\prime}.\text{collisions} \leftarrow \text{findCollisions}(R^{\prime})$
        \State Insert $R^{\prime}$ to OPEN
    \EndIf \label{newrootend}
    \State $\langle a_i, a_j, u, t \rangle$ \text{or} $\langle a_i, a_j, v, u, t \rangle \leftarrow \text{Earliest conflict in } P$ \label{conflictselect}
    \For{\textbf{each} agent $a_k$ \textbf{in} $\{a_i, a_j\}$ }\label{conflictstart}
        \State $Q.\text{root} \leftarrow$ \textit{false} \customcomment{$Q$ is a child node of $P$} 
        \State $Q.\text{sequencing} \leftarrow P.\text{sequencing}$
        \scriptsize
        \State $Q.\text{constraints} \leftarrow P.\text{constraints} \cup \{\langle a_k, u, t \rangle/\langle a_k, v, u, t \rangle\}$
        \State $Q.\text{paths}[a_k] \leftarrow \text{MLA}^*(a_k, Q.\text{sequencing}[a_k], Q.\text{constraints})$
        \small
        \If{$Q.\text{paths}[a_k]$ \textbf{is None}}
            \State \textbf{continue}
        \EndIf
        \State $Q.\text{cost} \leftarrow \text{flowtime}(Q.\text{paths})$
        \State Insert $Q$ to OPEN
    \EndFor\label{conflictend}
\EndWhile
\State \textbf{Return} No Solution
\normalsize
\end{algorithmic}
\end{algorithm}

\begin{figure*}[ht]
    \centering
    \begin{subfigure}[b]{0.20\textwidth}
        \centering
        \includegraphics[width=.9\textwidth]{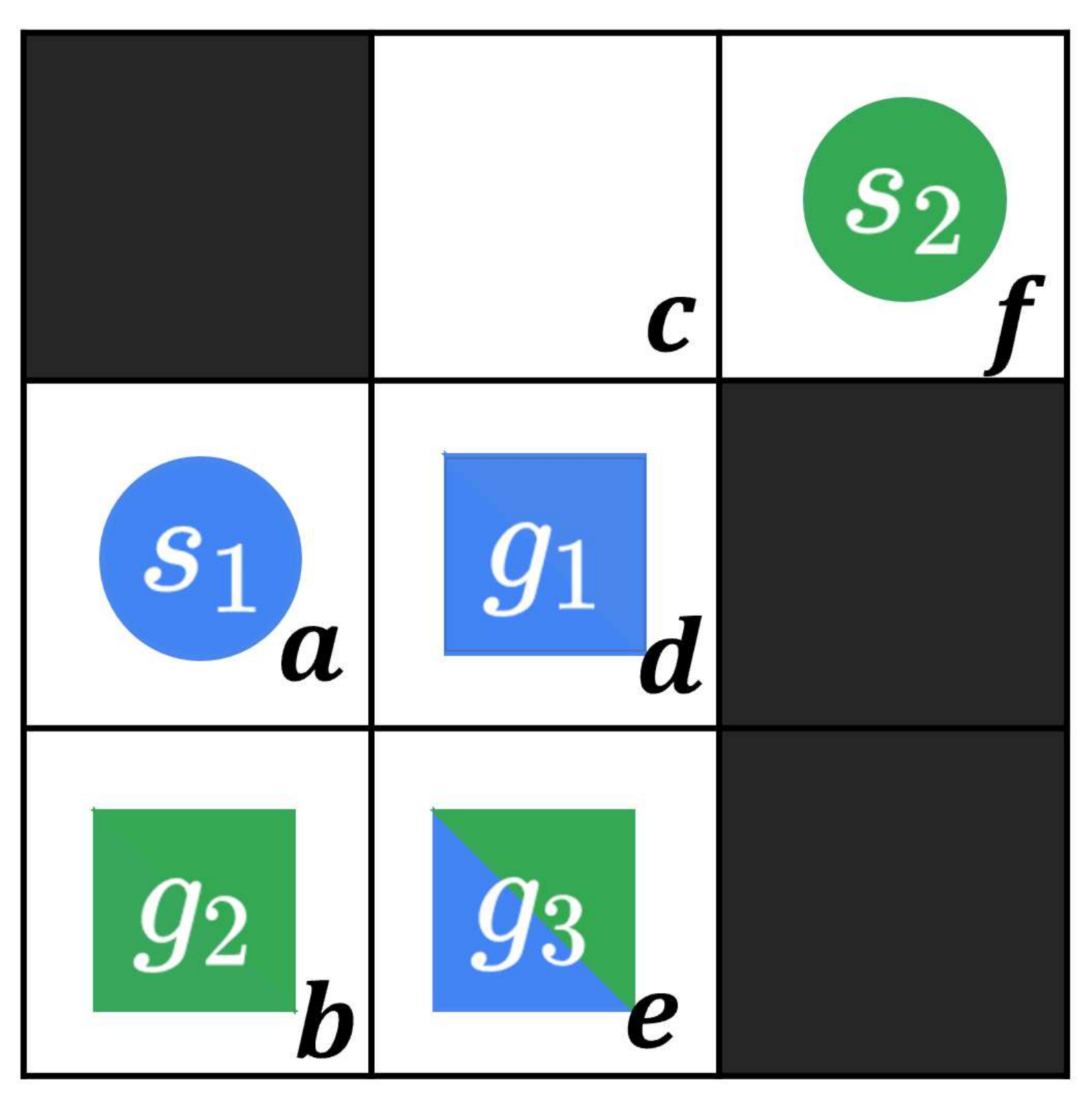}
        \caption{Start and task locations on a 3x3 grid map with agent-task compatibility constraints.}
        \label{fig:conflict forest map}
    \end{subfigure}
    \hfill 
    \begin{subfigure}[b]{0.75\textwidth}
        \centering
        \includegraphics[width=.9\textwidth]{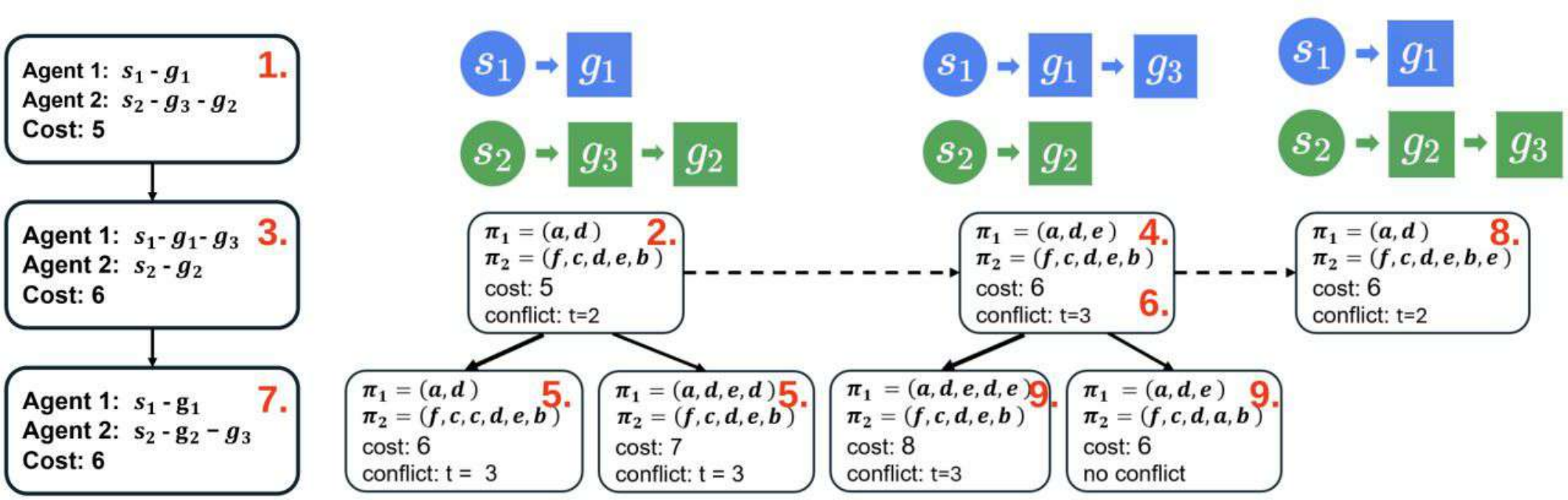}
        \caption{Best-first search on a conflict forest with the next-best task sequencing, where red numbers indicate the steps.}
        \label{fig:conflict forest}
    \end{subfigure}
    \caption{An example of conducting the best-first search on the conflict forest to find the optimal solution in CBS-TS.}
    \label{fig:example}
\end{figure*}

\subsection{Sequential Next Best Task Sequencing (MILP Formulation)}~\label{MILP}
 Let $V^{\prime}=\mathcal{S} \cup \mathcal{G}$ represent the union of initial and goal locations for all agents. Let $x_{ijk}$ be the binary decision variable, which equals 1 if agent $k$ traverses from location $i$ to $j$, and 0 otherwise. Since the task sequencing phase ignores potential conflicts, $c_{ij}$ in Cost matrix $C$ represents the travel cost between locations $i$ and $j$, which is the shortest distance obtained from A* planner. We define an auxiliary variable $y_{ik}$ to handle sub-tour elimination constraints. 
 
 The MILP formulation is as follows:
\small
\begin{equation}
{\operatorname{minimize}} \ \sum_{k \in \mathcal{A}}\sum_{i \in V^{\prime}} \sum_{j \in V^{\prime}} c_{i j} x_{i j k} \label{milp}
\end{equation}

Subject to
\begin{gather}
\sum_{i \in V^{\prime}} \sum_{k \in \mathcal{A}} x_{i j k}=1 \quad \forall j \in \mathcal{G} \label{forvg}\\
\sum_{j\in V^{\prime}} \sum_{k \in \mathcal{A}} x_{ijk}=1 \quad \forall i \in \mathcal{S} \label{forvd}\\
\sum_{i\in V^{\prime}} x_{ijk}-\sum_{i \in V^{\prime}} x_{jik}=0 \quad \forall
k \in \mathcal{A}, j \in \mathcal{G} \label{flowconserv}\\
y_{ik}-y_{jk}+\Tilde{M} x_{i j k} \leqslant \Tilde{M}-1 \quad  \forall (i,j) \in E,  k \in \mathcal{A} \label{subtour}\\
x_{ijk} \in\{0,1\} \quad \forall i \in V^{\prime}, j \in V^{\prime}, k \in \mathcal{A}\\
x_{i j k} \leq \delta_{k j} \quad \forall i \in V^{\prime}, j \in G, k \in \mathcal{A}\label{compatibility}
\end{gather}
\normalsize

Constraints~\eqref{forvg} and \eqref{forvd} ensure that each task location is visited exactly once by an agent, and each agent starts from its designated initial location. Constraints~\eqref{flowconserv} and \eqref{subtour} are the flow conservation and sub-tour elimination constraints, respectively, while the constraint~\eqref{compatibility} ensures that agent $i$ can only visit goal $j$ if they are compatible. 

In the default MILP formulation, agents would return to their start locations. To prevent this, we modify the formulation by treating start locations as dummy goals with zero travel costs. This ensures that optimal task sequencing is not influenced by return trips. During low-level pathfinding, the final segment of the route is discarded to enforce the non-return condition.

To obtain the next best task sequencing for the construction of the conflict forest, we impose additional constraints on the MILP to exclude previously found solutions. Two approaches are proposed. The first approach is a \textbf{\textit{direct constraint}}~\eqref{direct}, which exludes all the previous solutions simultaneously. Since task sequencing requires exactly one visit to all the goals and dummy goals (starts), the sum of corresponding decision variables $x_{ijk}$ equal to the total number of start locations and goal locations. The direct constraint ensures that the new solution differs from all previous solutions in at least one element—either by reassigning a task to a different agent or altering the sequencing of assigned tasks:

\begin{equation}
\sum_{k \in \mathcal{A}} \sum_{(i, j) \in P_k} x_{i j k} \leq |\mathcal{A}|+ |\mathcal{G}| -1, \quad \forall \mathcal{P} \in \mathcal{P}_{\text {set }}
\label{direct}
\end{equation}
where $\mathcal{P}_{\text{set}}$ is the set of previously found sequencing. And $P_k$ denote the set of visiting order $\left(i,j\right)$ of agent $k$ in a given sequencing $\mathcal{P}$, 

The second approach is \textbf{\textit{iterative constraints}} that excludes the previously found solutions one at a time. In each iteration, a single solution $\mathcal{P}=\left\{\mathcal{R}_1,\mathcal{R}_2,\ldots,\mathcal{R}_N\right\}$, where $\mathcal{R}_i$ is the visiting sequence for agent $i$, is excluded. We define $M$ as a non-empty subset of the solution space $\mathcal{Q}$. The iterative constraints are constructed as follows:
\begin{equation}
\begin{aligned}
M_1 & =\left\{\left(\overline{\mathcal{R}_1}\right)\right\} \\
M_2 & =\left\{\mathcal{R}_1 ;\left(\overline{\mathcal{R}_2}\right)\right\} \\
\quad & \ldots \\
M_n & =\left\{\mathcal{R}_1, \cdots, \mathcal{R}_{n-1} ;\left(\overline{\mathcal{R}_n}\right)\right\} \quad \text { for } n=1, \cdots, N .
\end{aligned}
\end{equation}
where $\overline{\mathcal{R}_n}$ and $\mathcal{R}_n$ represent enforcing and  prohibiting the task sequence of agent $n$ in solution $\mathcal{P}$. These subsets $M_n$ are mutually disjoint, and their union covers the entire solution space excluding $\mathcal{P}$~\cite{murty1968algorithm}. By solving the MILP within these subsets, we ensure a full exploration of the solution space while excluding $\mathcal{P}$. The mathematical formulations for banning and enforcing a specific route are as follows:

\begin{enumerate}
    \item \textbf{Ban a specific route $\overline{R_k} = \left\{v_{i_1}, v_{i_2}, \ldots, v_{i_n}\right\}$:}\\
    Limits the agent to traverse at most $n-2$ edges in this sequence.
    \begin{equation}
    \sum_{j=1}^{n-1} x_{i_j i_{j+1} k} \leq n-2
    \label{banconstraint}
    \end{equation}
    \item \textbf{Enforce a specific route $R_k = \left\{v_{i_1}, v_{i_2}, \ldots, v_{i_n}\right\}$:}\\
    Ensure that agent $k$ follows the exact sequence without deviation.
\end{enumerate}
\vspace{-0.1cm}
\begin{equation}
x_{i_j i_{j+1} k}=1 \quad \forall j \in\{1,2, \ldots, n-1\}
\label{enforceroute}
\end{equation}

The MILP can be solved with any off-the-shelf solver; in our experiments, we use Gurobi~\cite{gurobi}. In Algorithm~\ref{alg:highlevel}, $\text{MILP}(N,M,C,\Delta, \mathcal{P}_{\text{set}})$ represents the Gurobi MILP solver, which takes the number of agents $N$, the number of tasks $M$, the compatibility matrix $\Delta$, the cost matrix $C$ and the previously excluded sequencings $\mathcal{P}_{\text{set}}$ as input.

\begin{table*}[t]
    \centering
    \caption{Performance of four algorithms (CBSS, CBS-TS\textsubscript{i}, CBS-TS\textsubscript{d}, ECBS-TS) on 8$\times$8 maps with agent-task constraints}
    \setlength{\tabcolsep}{4pt} 
    \begin{tabular}{|c|c|c c c c|c c c c|c c c c|c c c c|}
        \hline
        \multirow{2}{*}{\textbf{N}} & \multirow{2}{*}{\textbf{M}} & \multicolumn{4}{c|}{\textbf{CBSS}} & \multicolumn{4}{c|}{\textbf{$\text{CBS-TS}_i$}} 
        & \multicolumn{4}{c|}{\textbf{$\text{CBS-TS}_d$}}
        & \multicolumn{4}{c|}{\textbf{ECBS-TS}}\\ 
        \cline{3-18}
          &  & \textbf{SR($\%$)} & \textbf{Cost} & $\bm{t} \left(s\right)$
         & \textbf{Op}$\left(s\right)$ & \textbf{SR}$\left(\%\right)$ & \textbf{Cost} & $\bm{t} \left(s\right)$
         & \textbf{Op}$\left(s\right)$ & \textbf{SR}$\left(\%\right)$ & \textbf{Cost} & $\bm{t} \left(s\right)$
         & \textbf{Op}$\left(s\right)$ & \textbf{SR}$\left(\%\right)$ & \textbf{Cost} & $\bm{t} \left(s\right)$
         & \textbf{Op$\left(10^{-5}s\right)$}\\
        \hline
        \multirow{3}{*}{5}  
          & 10  & 100 & 27.58 & 2.24 & 2.16 & 93.44 & 25.93 & 1.16 & 0.95 & 90.16 & 25.93 & 0.70 & 0.60 & 93.44 & 26.46 & 0.39 & 2.19\\
          & 20  & 96.61 & 40.62 & 13.99 & 13.81 & 91.53 & 39.53 & 4.02 & 3.29 & 91.53 & 39.53 & 6.16 & 5.47 & 94.92 & 40.51 & 0.87 & 2.19\\
          & 30  & 51.92 & 49.67 & 8.56 & 8.46 & 68.63 & 49.67 & 9.67 & 7.00 & 68.63 & 49.67 & 9.65 & 7.17 & 90.20 & 50.04 & 3.58 & 2.55\\
        \hline
        \multirow{3}{*}{10}  
          & 10 & 98.31 & 20.69 & 11.27 & 11.13 & 91.53 & 19.98 & 2.16 & 2.00 & 89.83 & 19.98 & 2.63 & 2.49 & 100 & 20.63 & 1.09 & 2.84\\
          & 20 & 28.00 & 30.14 & 8.39 & 8.31 & 72.00 & 30.14 & 10.14 & 9.56 & 72.00 & 30.14 & 9.93 & 9.36 & 94.00 & 30.64 & 2.78 & 2.91\\
          & 30 & 25.00 & 42.92 & 19.88 & 19.73 & 55.77 & 42.92 & 21.39 & 19.53 & 53.85 & 42.92 & 13.54 & 11.91 & 92.31 & 43.08 & 4.02 & 3.22\\
        \hline
        \multirow{3}{*}{20}  
          & 10 & 33.93 & 14.05 & 6.20 & 6.07 & 62.50 & 14.05 & 6.52 & 6.25 & 62.50 & 14.05 & 2.50 & 2.27 & 92.86 & 14.37 & 4.38 & 4.70\\
          & 20 & 24.07 & 24.66 & 24.59 & 24.38 & 35.19 & 24.66 & 9.13 & 8.35 & 37.04 & 24.66 & 3.80 & 3.05 & 77.78 & 24.92 & 7.41 & 4.80\\
          & 30 & 5.88 & 34.33 & 48.30 & 47.99 & 27.45 & 34.33 & 18.95 & 16.94 & 31.37 & 34.33 & 8.52 & 6.72 & 84.31 & 36.33 & 9.08 & 5.17\\
          \hline
    \end{tabular}
    \label{table2}
\end{table*}

\subsection{Low level of CBS-TS: Multi-label A*}~\label{lowlevel}
The Multi-Label A* (MLA*)~\cite{grenouilleau2019multi} algorithm extends the classical $A^*$ search method to optimally handle scenarios where an agent must visit a sequence of predefined task locations. Each state is represented as a tuple containing the agent’s current location, time, and task index. The task index, which tracks the current task, only updates when the agent reaches a task location, advancing to the next task in the sequence. This process continues until the agent reaches the final task.

MLA* uses an advanced heuristic that estimates the cost to reach the next task location, along with the cumulative cost of completing all remaining tasks. This heuristic, combined with the index-based state representation, ensures that MLA* finds not only a feasible path but also the optimal one, minimizing total travel time or cost.

\subsection{Example}
In Figure~\ref{fig:example}, we provide a step-by-step explanation of how CBS-TS operates using the same example presented in Figure~\ref{fig:concept}, with the red numbers indicating the sequence of steps. The algorithm first computes the best task sequencing (i.e. minimum-cost sequence) and corresponding paths while initially ignoring conflicts, forming the root node, which is added to the OPEN list (Steps 1-2). The node with the lowest cost is always selected for expansion. If the selected node is a root node, a new root node is first created by computing the next-best task sequencing before resolving conflicts (Steps 3-4). Then, the first conflict in the selected node is identified and resolved (Step 5). The process repeats: the lowest-cost node is selected (Step 6), and if it is another root node, a new root is generated before conflict resolution (Steps 7-8). Finally, conflict-free paths are obtained by resolving conflicts in the selected node (Step 9).
\subsection{Properties of CBS-TS}
\textbf{Theorem 1} - CBS-TS is complete.

\textbf{Proof.}
The completeness of CBS-TS relies on the following key points. First, CBS\cite{sharon2015conflict} is known to be complete for MAPF. MLA*\cite{grenouilleau2019multi} does not lose any feasible paths when navigating through a set of ordered goals. Thus, for a given task sequence, CBS-TS is guaranteed to find a conflict-free solution if one exists. Secondly, each time a root node is expanded, the next-best sequencing is computed until all possible sequencing have been explored. Since CBS-TS exhaustively explores both task sequencing (via MILP) and path planning (via CBS), it is guaranteed to find a solution if one exists.

\textbf{Theorem 2} - CBS-TS guarantees an optimal solution that minimizes the flowtime, provided such a solution exists.

\textbf{Proof.}
The optimality of CBS-TS stems from its systematic exploration of task sequences and conflict resolution. Each search tree in CBS-TS corresponds to a fixed task sequencing.  The cost of the root node of a tree serves as a lower bound for that tree because the MLA* computes paths without any constraints, and any conflict resolution by imposing constraints can only maintain or increase the cost. 

If the root node has conflict, CBS-TS does not immediately branch into child nodes to resolve them. Instead, it explores the next-best task sequencing, which becomes the root node of a new tree, thereby constructing a search forest. CBS-TS performs a best-first search over this search forest, the new sequencing will only be explored if it has a lower cost than current minimum cost. This guarantees that the cost of the found optimal solution is minimal among all explored task sequences and is no greater than the lower bounds of all unexplored root nodes, which represent the unexplored task sequences.
\begin{figure*}
    \centering
    \begin{subfigure}[b]{.3\linewidth}
         \centering
         \includegraphics[width=.92\linewidth]{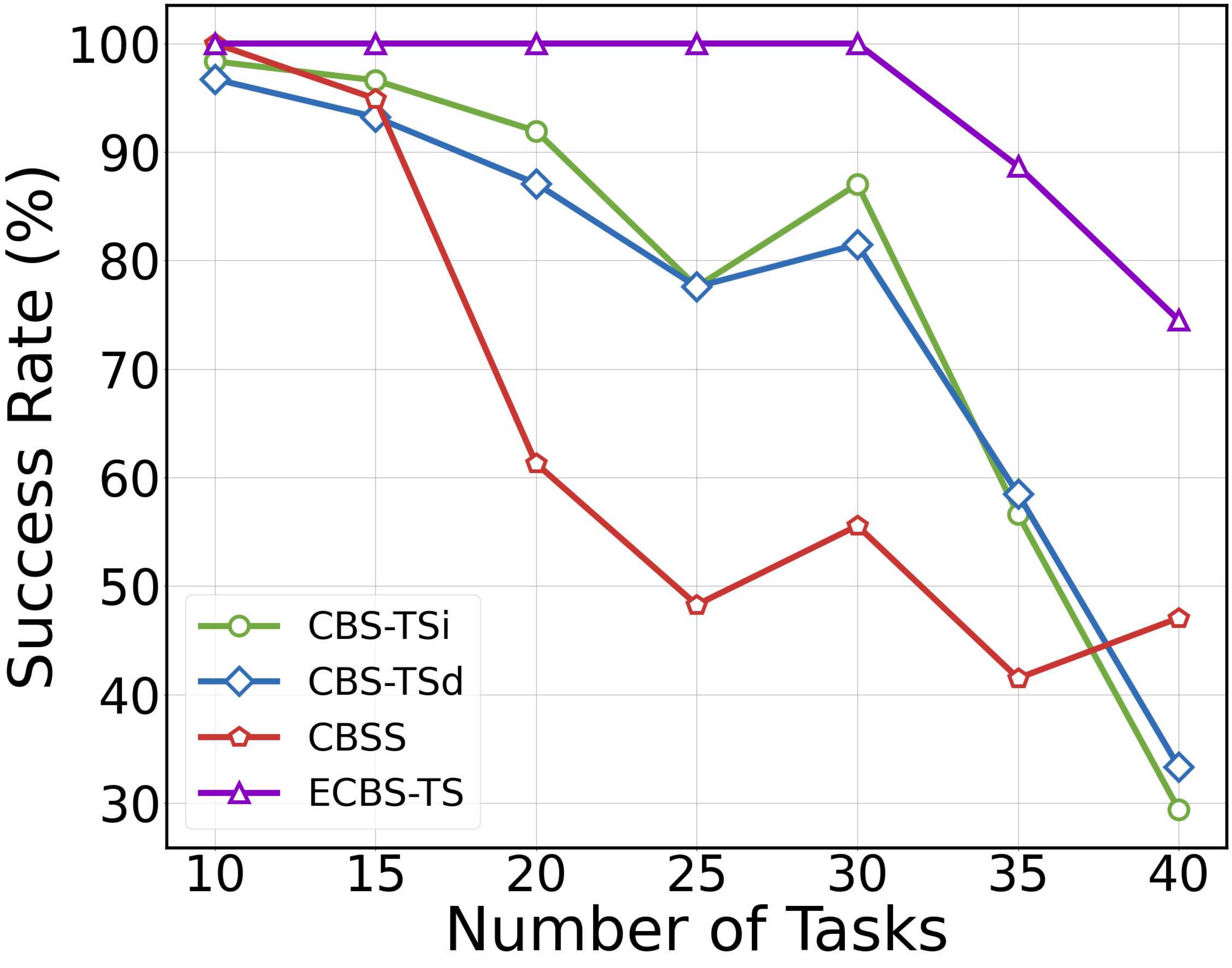}
         \caption{10 agents}
    \end{subfigure}
    \begin{subfigure}[b]{.3\linewidth}
         \centering
         \includegraphics[width=.92\linewidth]{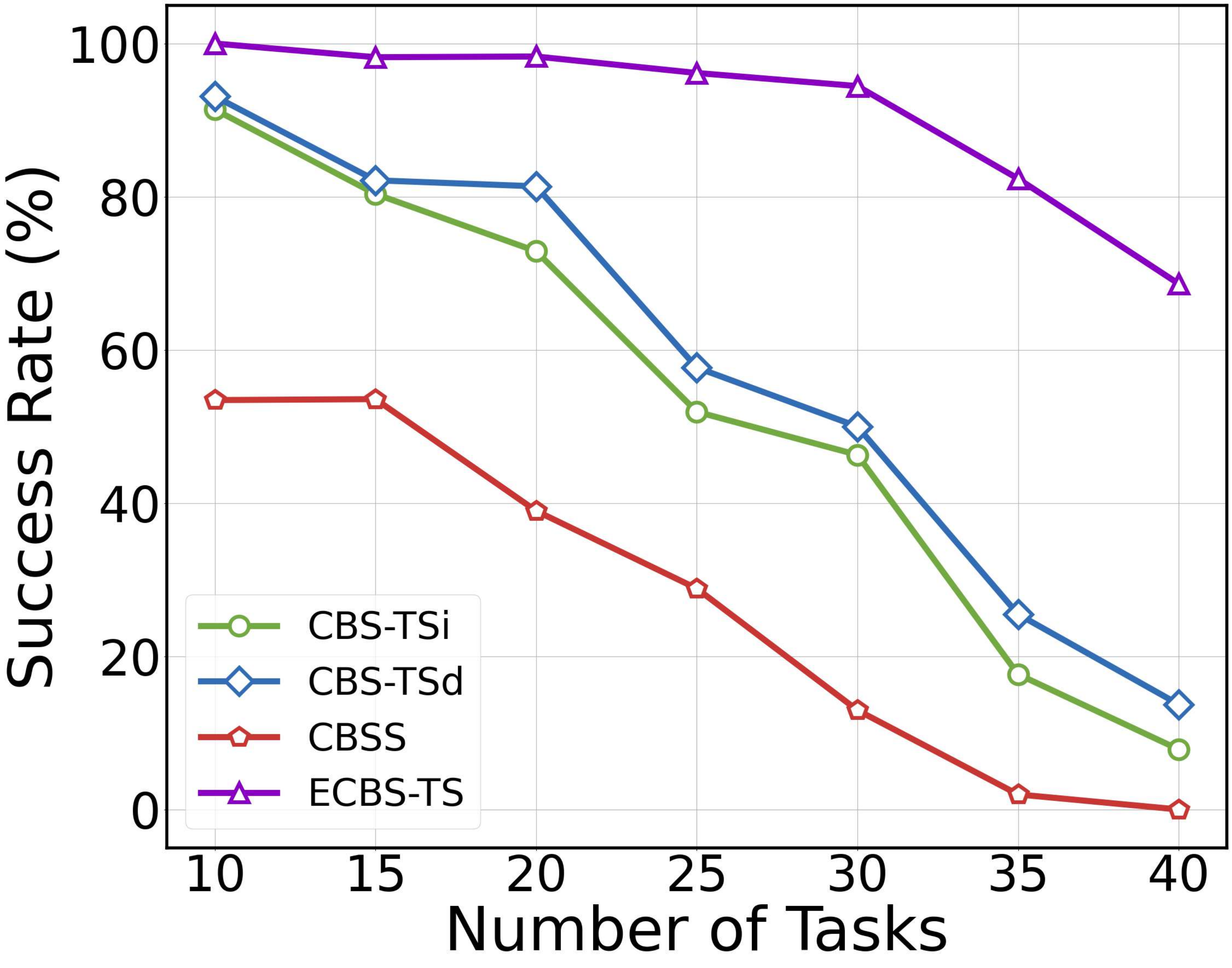}
         \caption{20 agents}
    \end{subfigure}
    \begin{subfigure}[b]{.3\linewidth}
         \centering
         \includegraphics[width=.92\linewidth]{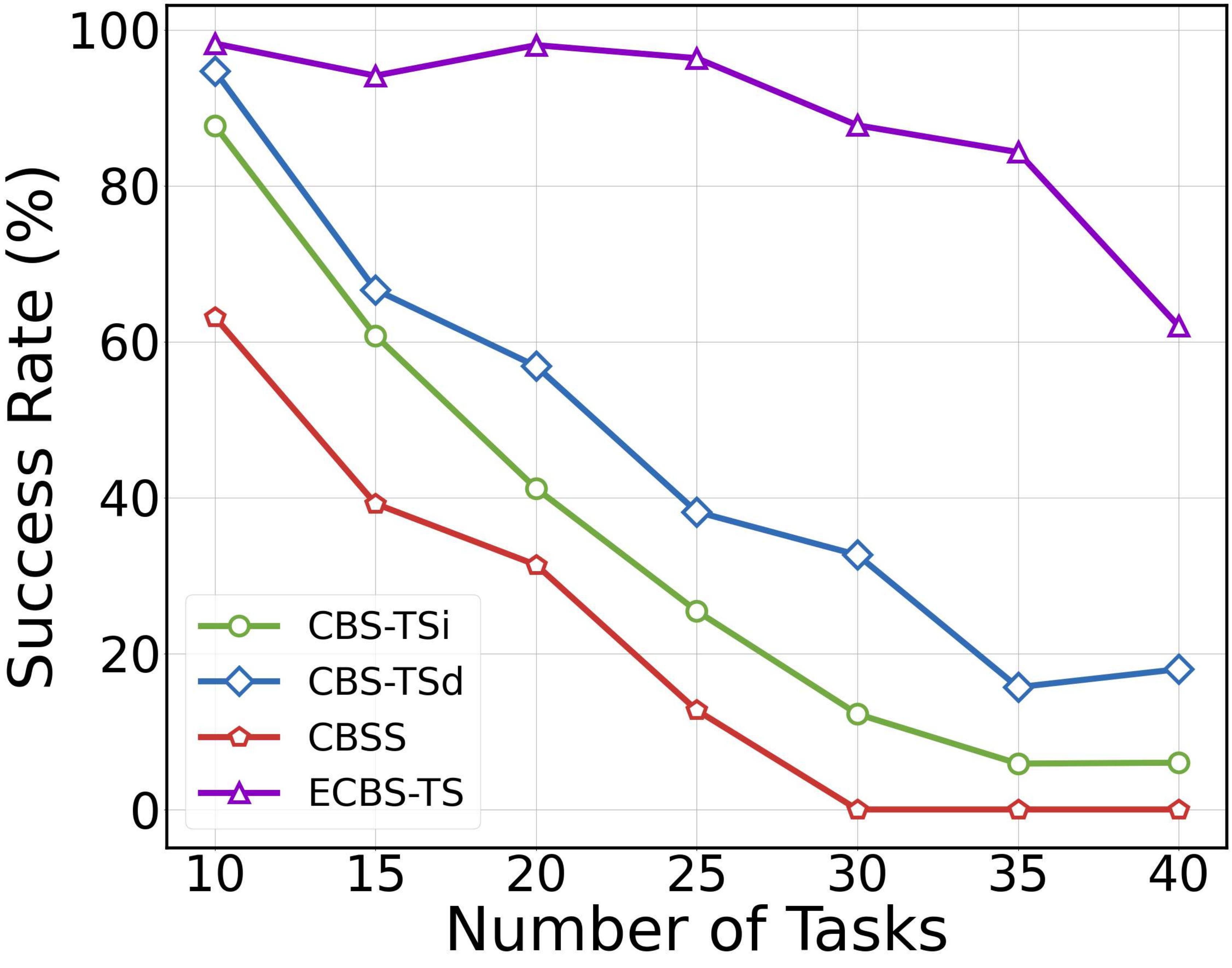}
         \caption{30 agents}
    \end{subfigure}
    \caption{Success rate with 60$s$ time limit.}
    \label{fig:successrate}
\end{figure*}

\begin{figure*}
    \centering
    \begin{subfigure}[b]{.3\linewidth}
         \centering
         \includegraphics[width=.92\linewidth]{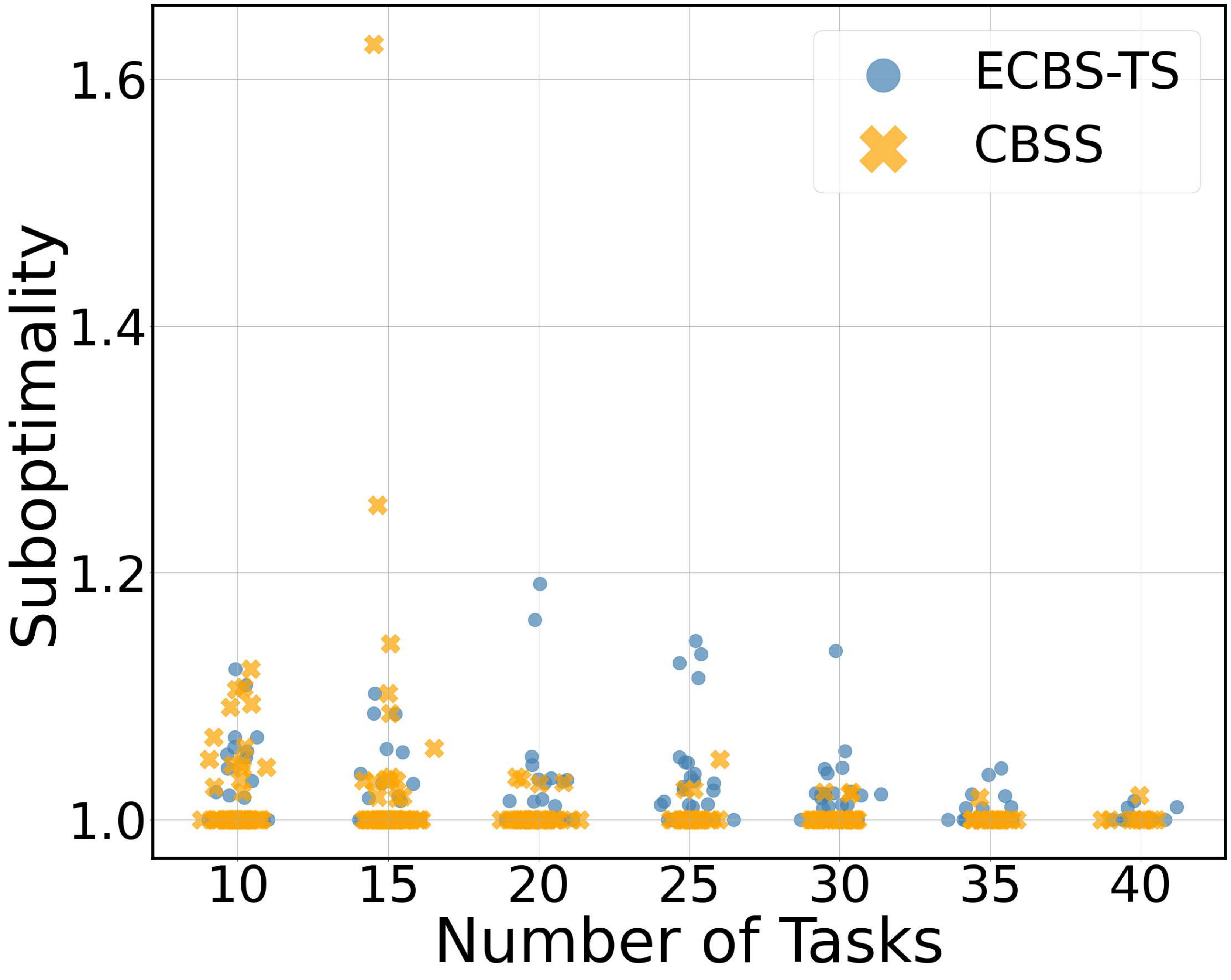}
         \caption{10 agents}
    \end{subfigure}
    \begin{subfigure}[b]{.3\linewidth}
         \centering
         \includegraphics[width=.92\linewidth]{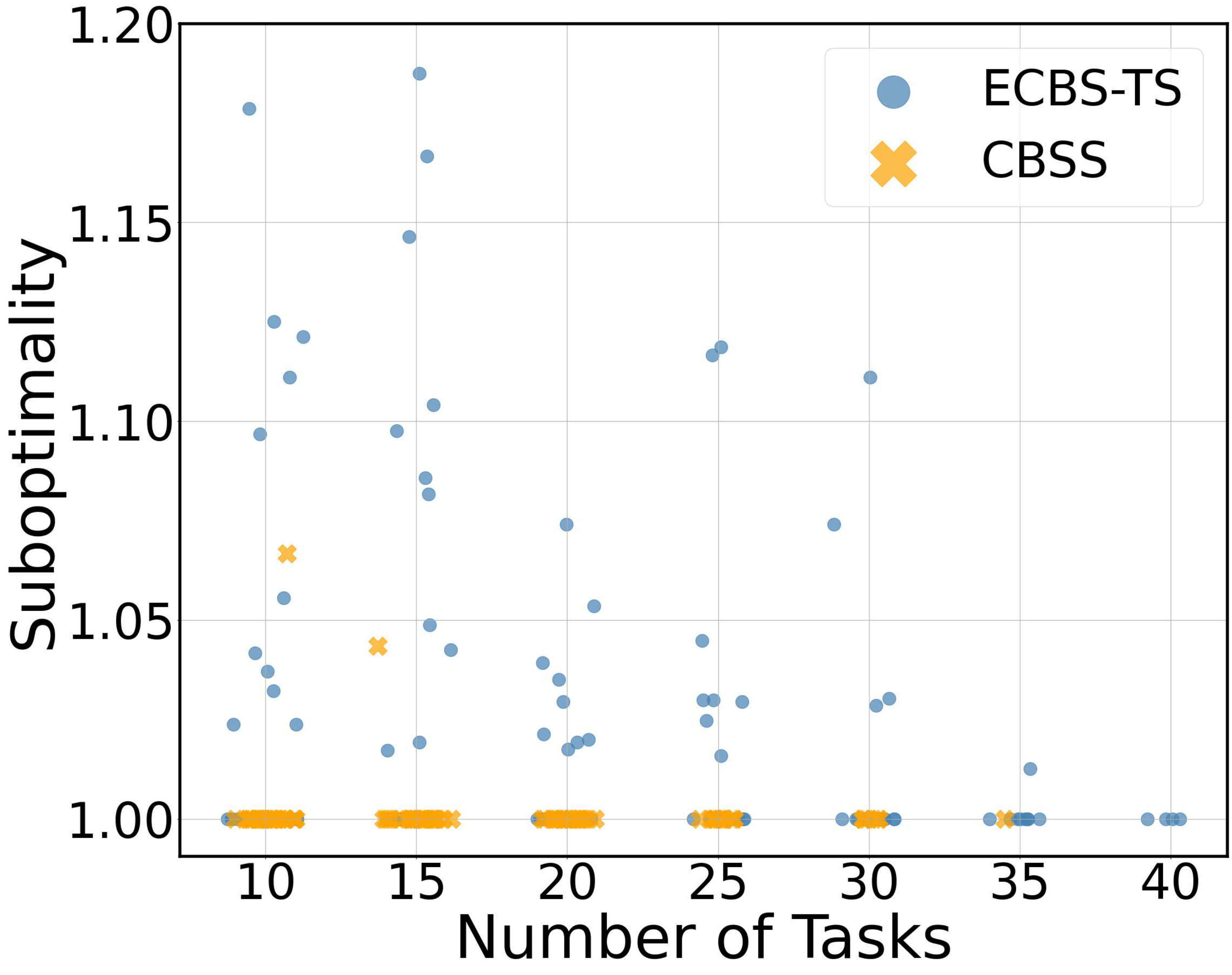}
         \caption{20 agents}
    \end{subfigure}
    \begin{subfigure}[b]{.3\linewidth}
         \centering
         \includegraphics[width=.92\linewidth]{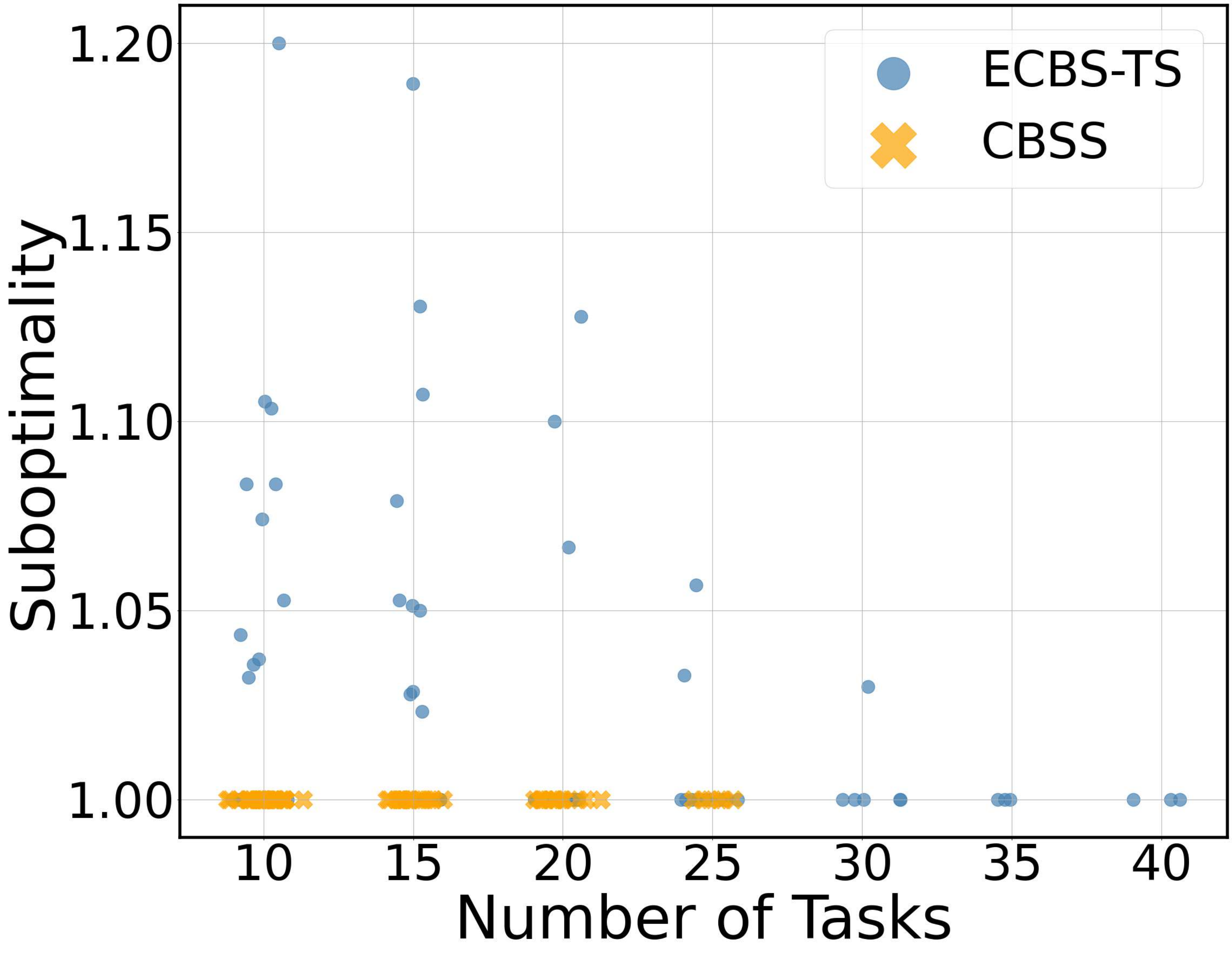}
         \caption{30 agents}
    \end{subfigure}
    \caption{Suboptimality ratio for CBSS and ECBS-TS. Only successful cases are included.}
    \label{fig:suboptimal}
\end{figure*}

\begin{figure*}
    \centering
    \begin{subfigure}[b]{.3\linewidth}
         \centering
         \includegraphics[width=.92\linewidth]{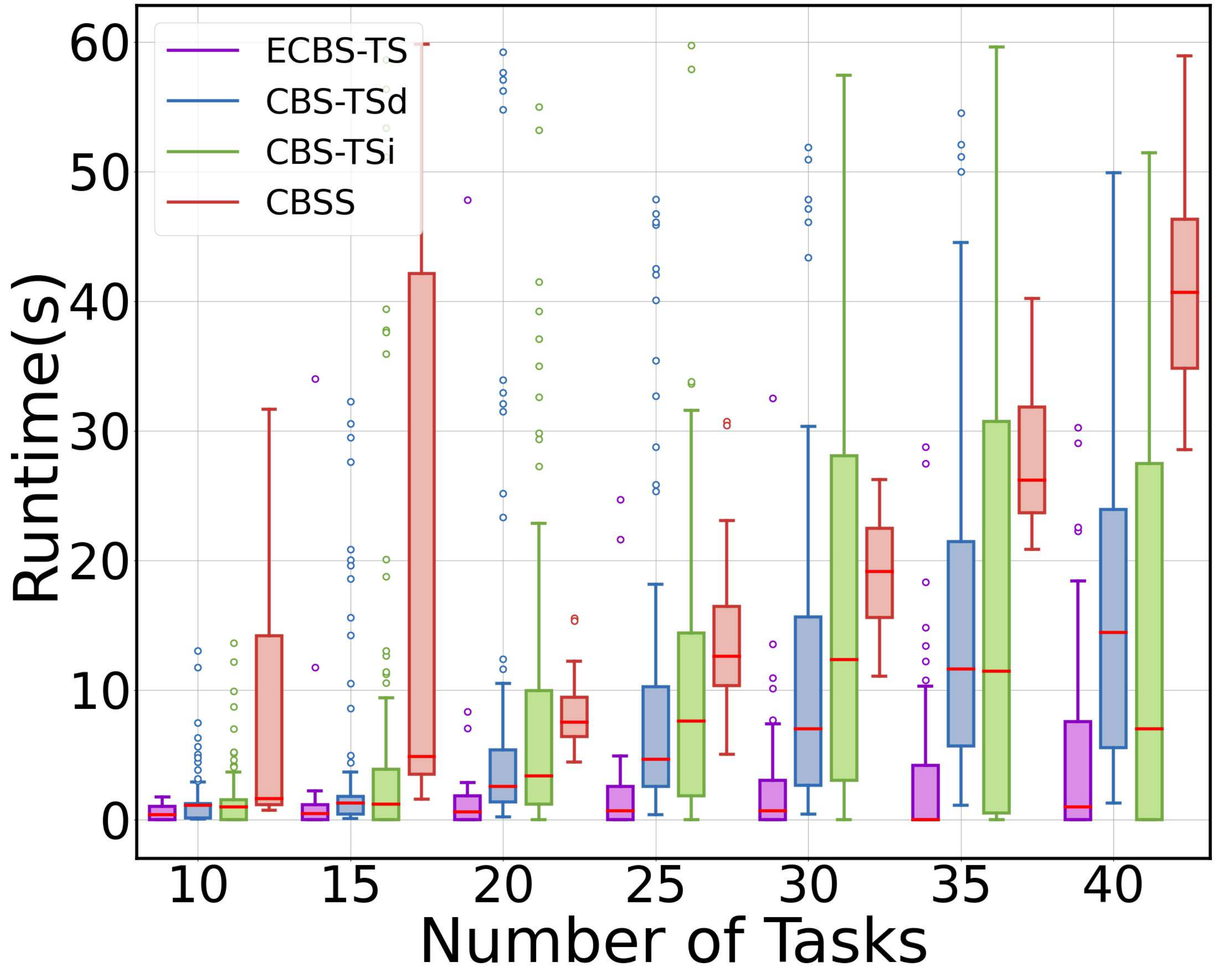}
         \caption{10 agents}
         \label{fig:exMap}
    \end{subfigure}
    \begin{subfigure}[b]{.3\linewidth}
         \centering
         \includegraphics[width=.92\linewidth]{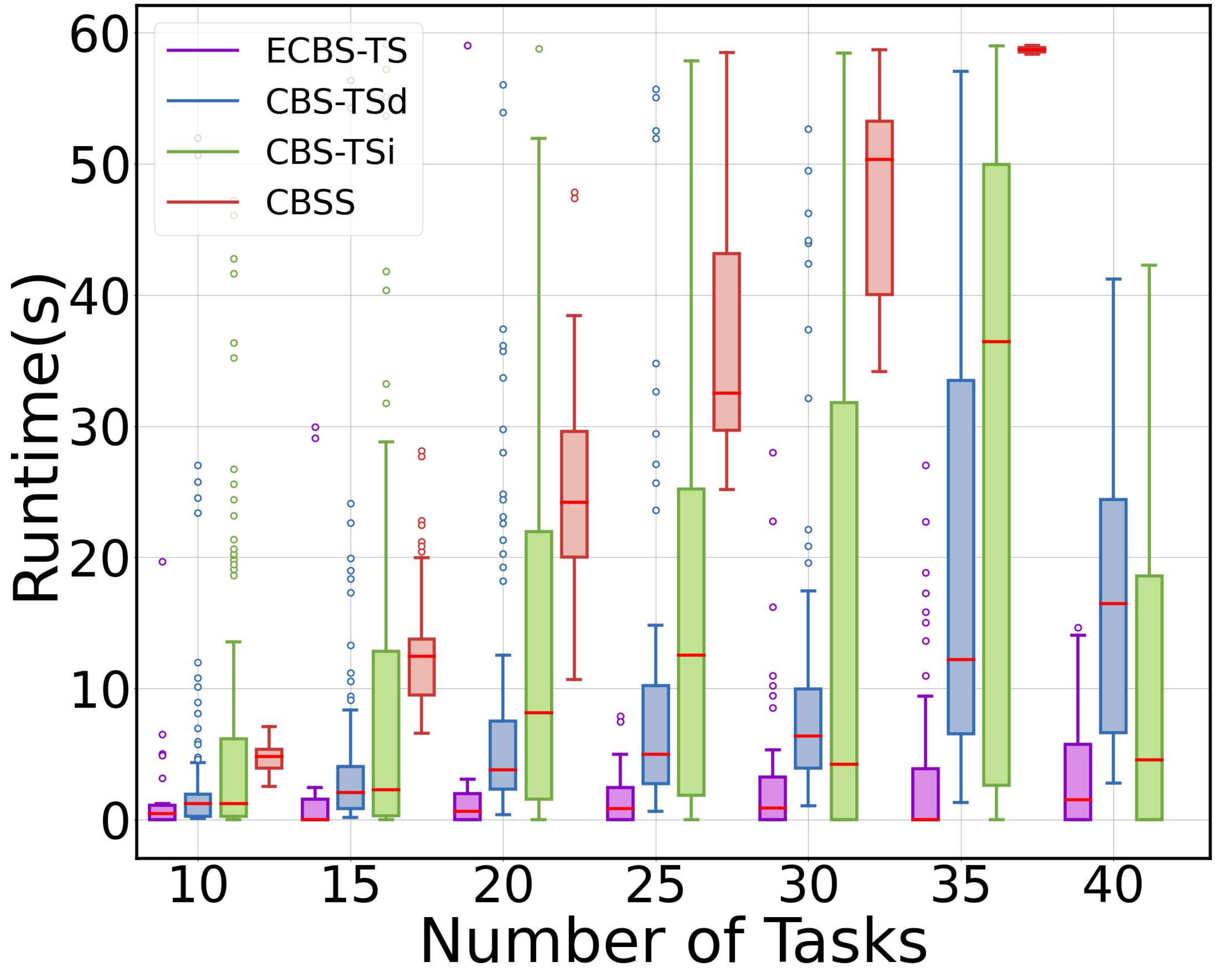}
         \caption{20 agents}
    \end{subfigure}
    \begin{subfigure}[b]{.3\linewidth}
         \centering
         \includegraphics[width=.92\linewidth]{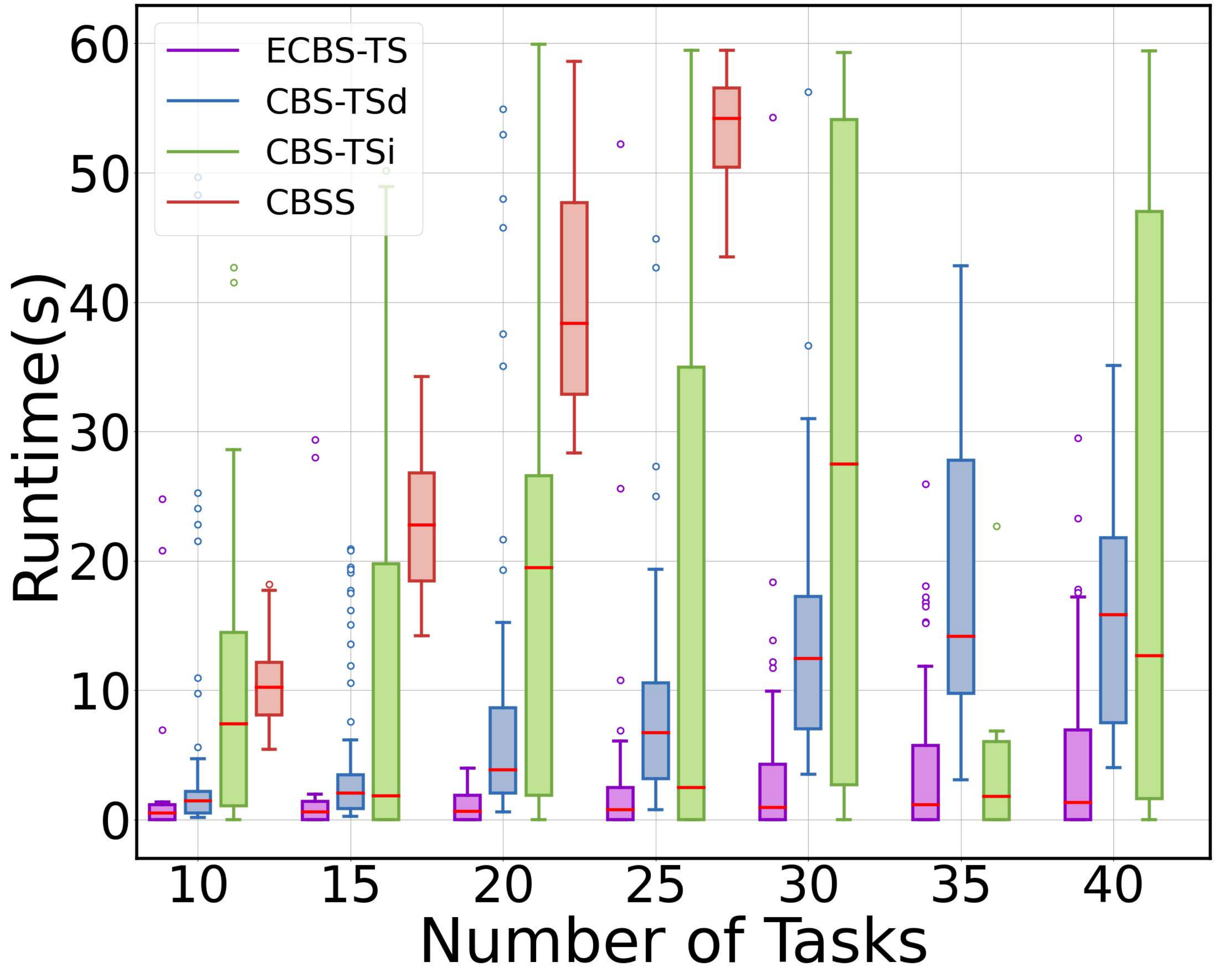}
         \caption{30 agents}
    \end{subfigure}
    \caption{Boxplot of runtime. Only successful cases are included.}
    \label{fig:boxplot}
\end{figure*}

\section{Experiments}
The proposed algorithms were implemented in C++ and tested on a 1.9GHz AMD Ryzen 7 Pro 5850U laptop with 16GB RAM, running Ubuntu 20.04 LTS. 
\subsection{Simulation}
We implement two variants of CBS-TS: CBS-TS\textsubscript{d}, which uses \textbf{\textit{direct constraint}}, and CBS-TS\textsubscript{i}, which uses \textbf{\textit{iterative constraints}}. Additionally, we introduce ECBS-TS, a bounded-suboptimal variant of CBS-TS\textsubscript{d}. ECBS-TS applies the same principles to CBS-TS\textsubscript{d} as ECBS~\cite{barer2014suboptimal} does to CBS, incorporating focal search at both the high and low levels. Specifically, ECBS-TS introduces a suboptimality factor $\omega$ to balance solution quality and computation time. By maintaining a FOCAL list, which contains nodes whose $f$-values are within $\omega$ times the minimum $f$-value in the OPEN list, ECBS-TS enables faster identification of suboptimal solutions. The solution is guaranteed to be at most $\omega$ times worse than the optimal solution. In our implementation, we set $\omega=1.2$.

We compare these methods against CBSS~\cite{ren2023cbss}, which addresses a similar problem but requires a subset of tasks—equal to the number of agents—to serve as destinations where agents must terminate. To adapt CBSS for TSPF problem and ensure a fair comparison with CBS-TS, we modify CBSS as follows: (1) We use the same task set as CBS-TS, ensuring that all tasks are visited; (2) We introduce dummy destinations and set the cost from any location to these destinations to 0, effectively removing termination constraints. This allows agents to move freely without being assigned fixed destinations, making CBSS equivalent to CBS-TS in this setting. It should be noted that while CBSS claims optimality, its public implementation uses a heuristic TSP solver that does not guarantee an optimal solution\footnote{\url{https://github.com/rap-lab-org/public_pymcpf}}.

We conduct experiments on randomly generated 8$\times$8 four-connected grids with 20\% obstacles. The scenarios are defined by varying numbers of agents $N \in \left\{5, 10, 20\right\}$ and tasks $M \in \left\{10, 20, 30\right\}$. The start locations of agents and task locations are unique and obstacle-free. Agents and tasks are assigned types to incorporate agent-task compatibility constraints: two types of agents (Agent $\alpha$ and Agent $\beta$) and three types of tasks (Task $A$, Task $B$, and Task $C$). Agent $\alpha$ can handle Task $A$ and Task $C$; while agent $\beta$ could execute Task $B$ and Task $C$. For each setting, we run 60 instances with a time limit of 60 seconds for all algorithms.

Table~\ref{table2} presents the performance of CBSS, CBS-TS\textsubscript{i}, CBS-TS\textsubscript{d}, and ECBS-TS on 8x8 maps.  The success rate (\textbf{SR}), represents the percentage of instances, out of a total of 60, where a solution is found within the time limit. $\bm{t}$ denotes the total computational time, and \textbf{Op} refers to the optimization time for task sequencing, with MILP for CBS-TS\textsubscript{i}, CBS-TS\textsubscript{d} and ECBS-TS, while mTSP for CBSS. \textbf{Cost} represents the average flowtime. For fairness, Cost, Op, and $t$ are compared only for instances where all algorithms succeed.

As $N$ and $M$ increase, the success rates of all algorithms decline. ECBS-TS, a bounded-suboptimal method, consistently achieves the highest success rates, remaining above 80\% even for 20 agents and 30 tasks. In contrast, CBSS experiences the steepest decline, dropping from 100\% in small instances to 5.88\% in the most complex case, highlighting its struggle to scale. CBS-TS\textsubscript{i} and CBS-TS\textsubscript{d} outperform CBSS, particularly in large instances, where they maintain significantly higher success rates. Runtime $t$ increases with $N$ and $M$ for all algorithms. For CBS-TS\textsubscript{i}, CBS-TS\textsubscript{d} and CBSS, the majority of the runtime is consumed by optimization. In contrast, ECBS-TS spends minimal time on optimization, with most of its runtime dedicated to conflict resolution. In terms of solution quality, CBS-TS\textsubscript{i} and CBS-TS\textsubscript{d} yield the lowest, as they are optimal methods. CBSS demonstrates near-optimal behavior, with costs slightly higher than optimal in smaller instances but converging to optimality as the problem scale increases. ECBS-TS presents the highest cost, resulting from the trade-off between optimality and efficiency.

To assess scalability, we also analyze the algorithms on a larger 16$\times$16 map with $N \in \left\{10, 20, 30\right\}$ and $M \in \left\{10,15,20,25,30,35,40\right\}$. The performance trends on 16$\times$16 maps, illustrated in Figures \ref{fig:successrate}, \ref{fig:suboptimal}, and \ref{fig:boxplot}, align closely with those on 8$\times$8 maps.

Figure~\ref{fig:successrate} illustrates the success rates of all the methods. ECBS-TS consistently achieves the highest success rate. CBS-TS\textsubscript{i} and CBS-TS\textsubscript{d}, consistently outperform CBSS, except in some 10 agents cases. Between the two, CBS-TS\textsubscript{i} achieves a slightly higher success rate than CBS-TS\textsubscript{d} for 10 agents. However, as the agent count grows, $\text{CBS-TS}_d$ scales better, indicating that solving a more constrained MILP is more efficient than solving a less constrained MILP $N$ times when $N$ is large.

Figure~\ref{fig:suboptimal} presents the suboptimality ratio, defined as the flowtime of a suboptimal method divided by the flowtime of the optimal methods (CBS-TS\textsubscript{i}/CBS-TS\textsubscript{d}). ECBS-TS adheres to its 1.2 suboptimality bound, while CBSS demonstrates near-optimal behavior without a formal bound. Although most CBSS instances achieve suboptimality ratios below 1.2, there is one outlier exceeding 1.6 for 10 agents and 15 tasks. Notably, the scatter plot reveals a sparser distribution of points for CBSS suboptimal ratio in larger problem instances, with some cases disappearing entirely, reflecting the declining success rates of CBSS as the problem scale grows.

Figure~\ref{fig:boxplot} shows the runtime distributions across all algorithms for varying numbers of agents and tasks. ECBS-TS consistently achieves the lowest median runtime, followed by CBS-TS methods, while CBSS exhibits the highest median runtime. CBS-TS\textsubscript{i} exhibits higher variability than CBS-TS\textsubscript{d}, occasionally achieving lower quartiles. This variability stems from CBS-TS\textsubscript{i} partitions the solution space into $N$ subsets when searching for the next-best sequencing. Since the solution may reside in any of these subsets, instances where the solution falls into an earlier subset are solved more quickly, while others may take longer. However, in general, CBS-TS\textsubscript{i} requires more computation time than CBS-TS\textsubscript{d}, especially as the number of agents and tasks increases. In some challenging cases (e.g., 20 agents with 40 tasks or 30 agents with 35 tasks), CBS-TS\textsubscript{i} appears to have a lower runtime than CBS-TS\textsubscript{d}. However, this is due to its lower success rate—fewer successful instances mean only computationally easier cases are recorded, skewing the results.

\subsection{Physical Experiment}
\begin{figure}[!htbp]
    \centering
    \begin{subfigure}[b]{.48\linewidth}
         \centering \includegraphics[width=.9\linewidth]{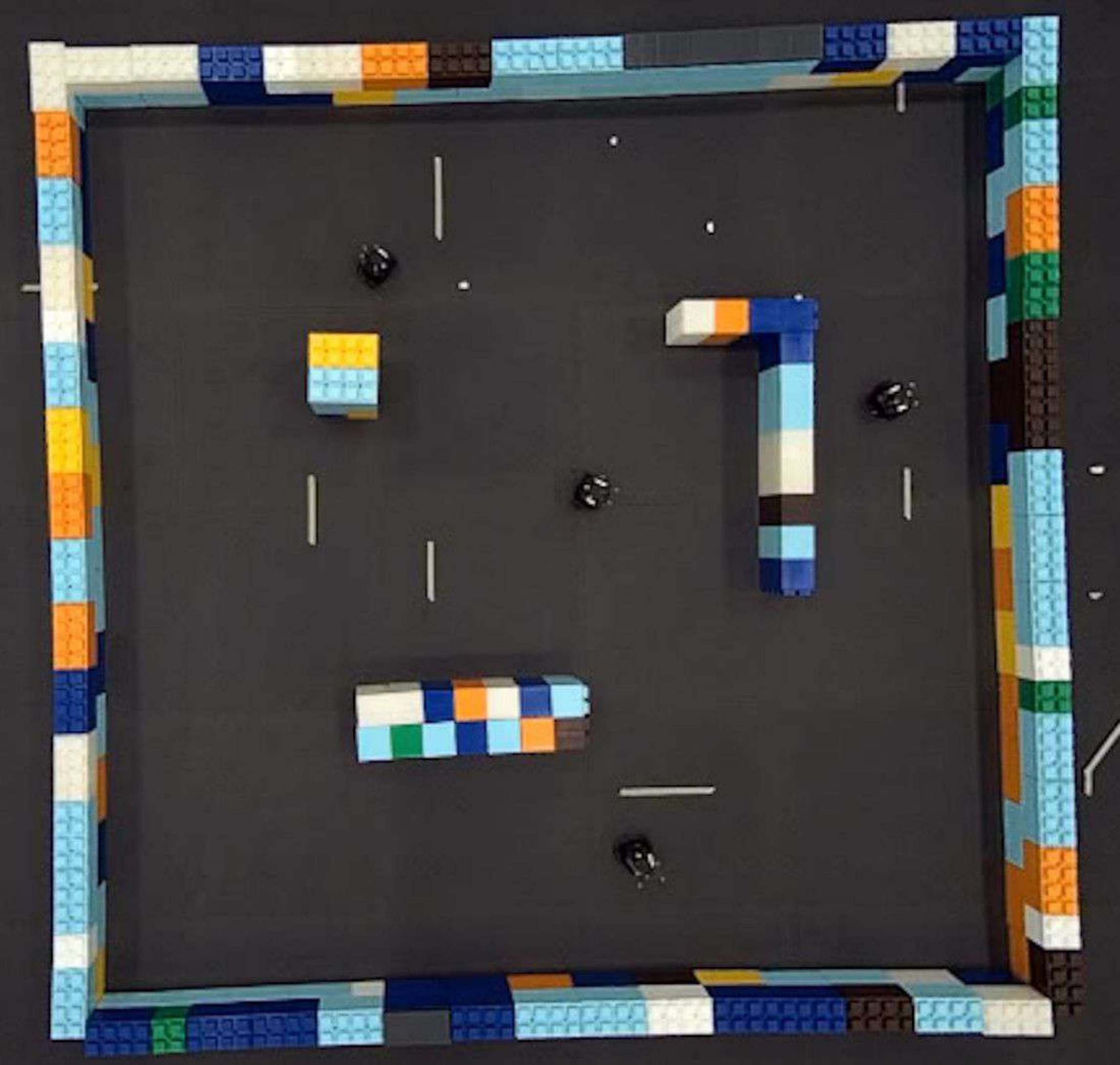}
         \caption{}
         \label{fig:sub1}
    \end{subfigure}
    \begin{subfigure}[b]{.48\linewidth}
         \centering
         \includegraphics[width=.9\linewidth]{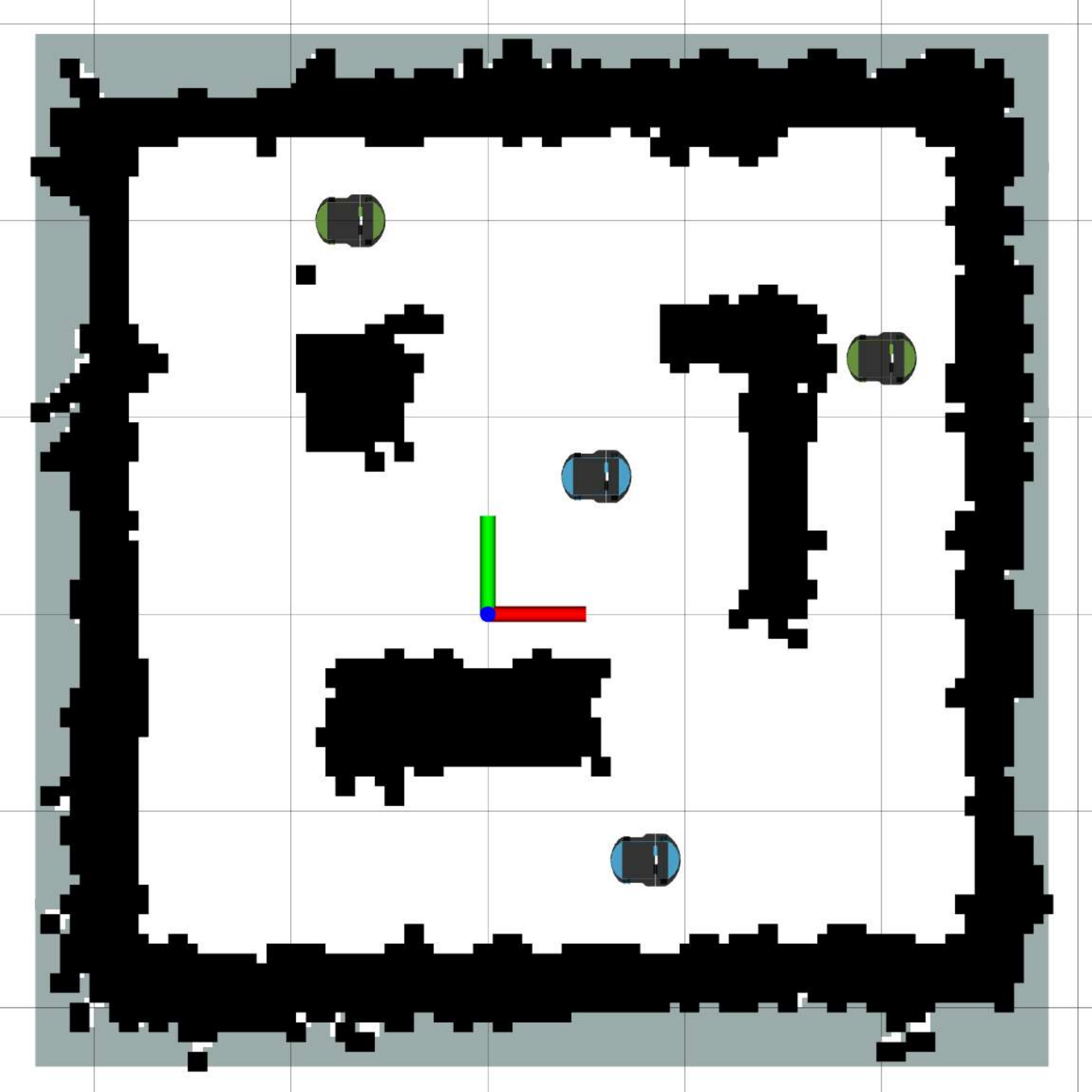}
         \caption{}
         \label{fig:sub2}
    \end{subfigure}
    
    \begin{subfigure}[b]{.48\linewidth}
         \centering
         \includegraphics[width=.9\linewidth]{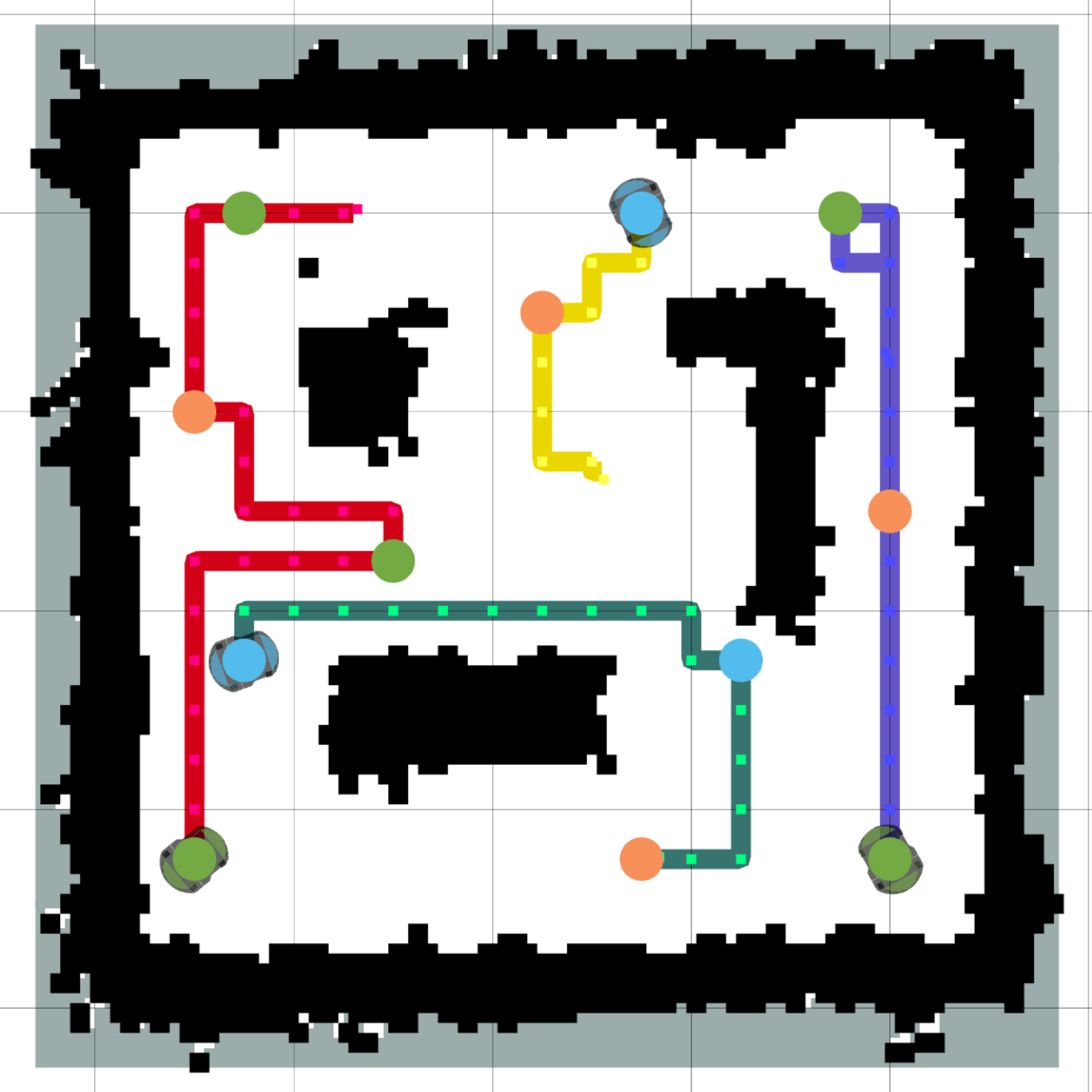}
         \caption{}
         \label{fig:sub3}
    \end{subfigure}
    \begin{subfigure}[b]{.48\linewidth}
         \centering
         \includegraphics[width=.9\linewidth]{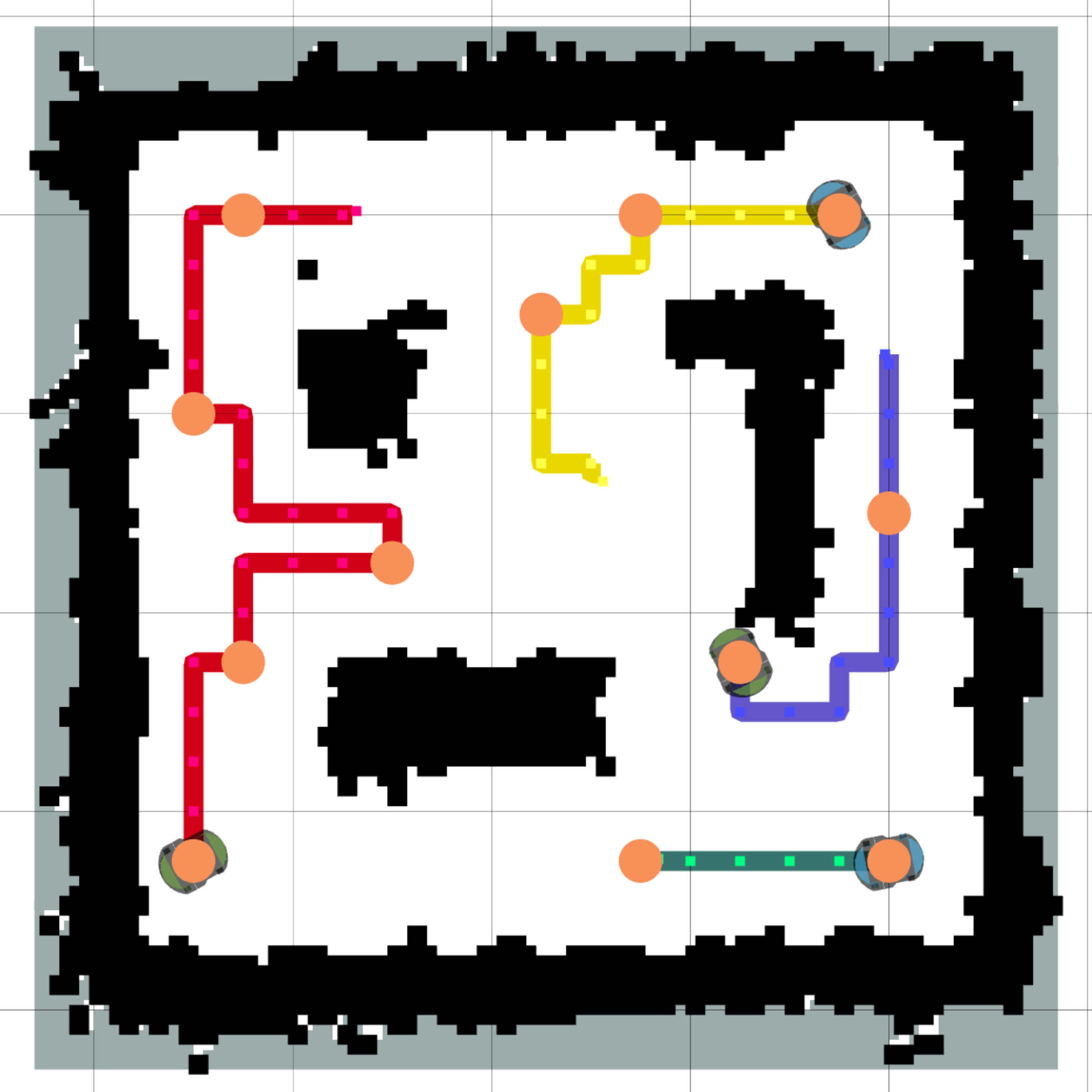}
         \caption{}
         \label{fig:sub4}
    \end{subfigure}
    \caption{Experimental setup and path visualization in Rviz. (a) Top view of the real-world environment. (b) Initial robot positions in Rviz: type $\alpha$ agents (blue) and type $\beta$ agents (green). (c) and (d) Planned paths and generated trajectories under different task-type distributions. Paths are represented by colored cubic markers, while trajectories are shown as colored lines. Tasks are depicted as colored circles: Task A (blue) is exclusive to type $\alpha$ agents, Task B (green) is exclusive to type $\beta$ agents, and Task C (orange) can be completed by either agent type.}
    \label{fig:t=30}
\end{figure}

We conducted real-world experiments with Turtlebot3 Burger robots in an indoor testing area of approximately 4.5m $\times$ 4.5m. A grid map with 0.25m resolution of the environment is built with \textit{octomap\_server}\cite{hornung2013octomap}. Figure~\ref{fig:sub1} illustrates the real-world setup, where four Turtlebot3 Burger are placed at 4 start locations. Two robots at $[0.56,0.66]$ and $[0.84, -1.24]$ are classified as type $\alpha$, while the other two at $[1.98,1.30]$ and $[-0.68,2.02]$ are classified as type $\beta$. Figure~\ref{fig:sub2} presents the start locations of the robots in Rviz. 

Figure~\ref{fig:sub3} and \ref{fig:sub4} compare the sequencing and trajectories generated for the agents under different task-type distributions, using the same 12 task locations but varying their task types. This highlights the impact of agent-task compatibility constraints on TSPF problems. 

We used CBS-TS\textsubscript{d} to determine the optimal paths for each robot to visit its assigned tasks. Based on these paths, a Model Predictive Controller (MPC) in \cite{li2020efficient} was employed to time-parameterized trajectories. The MPC assigns a time interval of 1s for each waypoint in the CBS-TS\textsubscript{d} paths. Given the grid resolution of 0.25m, this results in an expected speed of 0.25m/s. The MPC incorporates a robot collision model with a radius of $R=0.15$m, a maximum velocity of $v^{max}=1$m/s, and a maximum angular velocity of $\omega^{max}=1$rad/s. 

\section{Conclusion and Future Work}
In this paper, we introduced CBS-TS, an optimal and complete algorithm for solving the collaborative multi-agent task assignment, sequencing and pathfinding (TSPF) problem, incorporating agent-task assignment constraints and ensuring full task coverage. Through extensive simulations, CBS-TS not only guarantees optimality but also demonstrates better efficiency compared to the near-optimal implementation of the baseline method, CBSS. Real-world experiments validate the practical applicability of our framework, demonstrating its effectiveness in real-world scenarios.

For future work, we plan to incorporate nonlinear optimization techniques to smooth trajectories, ensuring dynamic feasibility for non-holonomic agents. Additionally, we aim to evaluate the robustness and scalability of CBS-TS in denser environments with more agents. This includes testing in constrained environments, such as narrow corridors where only one agent can pass at a time, as well as scenarios requiring agents to reposition after completing all tasks to allow others to proceed.

\bibliographystyle{IEEEtran}
\bibliography{IEEEabrv,references}
\end{document}